%% file: ms.tex
 \documentclass[preprint2]{aastex}
\usepackage{amsmath}
\usepackage{xspace}
\usepackage{multirow}
\usepackage{floatpag}
\usepackage{lscape}
\usepackage{rotating}
\usepackage{textcomp}
\newcommand{\myemail}{jdo@astro.unam.mx}
    \def\kms{\ensuremath{\, \rm{km \, s^{-1}}}}
    \def\ergs{\ensuremath{\, \rm{erg \, s^{-1}}}}
    \def\ergscm{\ensuremath{\, \rm{erg \, s^{-1} \, cm^{-2}}}}
    \def\ergscma{\ensuremath{\, \rm{erg \, s^{-1} \, cm^{-2} \, \text{{\AA}}^{-1}}}}
    \def\micron{\ensuremath{\mu\rm{m}}\xspace}
    \def\lolim{\ensuremath{10^{41}}} % \ergs
    \def\uplim{\ensuremath{10^{44}}} % \ergs
    \def\la{\ensuremath{\rm{Ly\alpha}}\xspace}
    \def\nsimul{5000\xspace}
    \def\obsdate{2010 September 8\xspace}
    \def\otf{\mbox{OSIRIS-TF}\xspace}
    \def\pa{\phantom{0}}
    
    \newcommand{\avg}[1]{\left< #1 \right>}
    \def\sun{\ensuremath{_{\rm{\odot}}}}
    \def\cluster{MS\,2053.7-0449\xspace}
    \newcommand{\tn}[1]{\tablenotemark{#1}}
    \newcommand{\mr}[1]{\multirow{2}{*}{#1}}
    \newcommand{\mc}{\multicolumn{2}{c}}       
    \newcommand{\mca}{\multicolumn{1}{c}}   
     \def\oii{[\ion{O}{2}]}
    \def\laea{\ensuremath{\mathbb{A}}a\xspace}
    \def\laeb{\ensuremath{\mathbb{A}}b\xspace}
    \def\laec{\ensuremath{\mathbb{A}}c\xspace}
    \def\dlla{\ensuremath{\mathbb{B}}a\xspace}
    \def\dllb{\ensuremath{\mathbb{B}}b\xspace}
    \def\lbga{\ensuremath{\mathbb{C}}a\xspace}
    \def\lbgb{\ensuremath{\mathbb{C}}b\xspace}
\slugcomment{Submitted to A\&A.}

\shorttitle{Filter Induced Bias}
\shortauthors{de Diego et al.}

\begin{document}

 \thispagestyle{empty}

\title{Filter Induced Bias in \la Emitter Surveys: \\
        A Comparison Between Standard and Tunable Filters. \\
        Gran Telescopio Canarias Preliminary Results}
    % Vedic Lyman $\alpha$-logy

\author{J. A. de Diego\altaffilmark{1} and M. A. De Leo}
\affil{Instituto de Astronom\'{\i}a;
Universidad Nacional Aut\'{o}noma de M\'{e}xico \\
Avenida Universidad 3000; Ciudad Universitaria; C.P. 04510;
Distrito Federal; M\'{e}xico
}
\email{\myemail}

\author{J. Cepa and A. Bongiovanni}
\affil{Instituto de Astrof\'{\i}sica de Canarias,
% V\'{\i}a L\'{a}ctea s/n,
E-38205 La Laguna, Tenerife, Spain \\ \& \\ Departamento de Astrof\'{\i}sica, Universidad de La Laguna, E-38206 La Laguna, Tenerife, Spain
}

\author{T. Verdugo}
\affil{Centro de Investigaciones de Astronom\'{\i}a (CIDA),
Apartado Postal 264, M\'{e}rida 5101-A, Venezuela
}

\author{M. S\'{a}nchez-Portal\altaffilmark{2}}
\affil{Herschel Science Centre (HSC), European Space Agency Centre (ESAC)/INSA, \\ Villanueva de la Ca\~{n}ada, Madrid, Spain}

\and

\author{J. I. Gonz\'{a}lez-Serrano}
\affil{Instituto de F\'{\i}sica de Cantabria (CSIC-Universidad de Cantabria), E-39005 Santander, Spain}

\altaffiltext{1}{Visiting Astronomer, Instituto de Astrof\'{\i}sica de Canarias.}
\altaffiltext{2}{Asociaci\'{o}n ASPID, Apartado de Correos 412, La Laguna, Tenerife, Spain}

\begin{abstract}
    % Background | Aim | Problem | What this paper does
    \la emitter (LAE) surveys have successfully used the excess in a narrow-band filter compared to a nearby broad-band image to find candidates. %
    However, the odd spectral energy distribution (SED) of LAEs combined with the instrumental profile have important effects on the properties of the candidate samples extracted from these surveys. %
    We investigate the effect of the bandpass width and the transmission profile of the narrow-band filters used for extracting LAE candidates at redshifts $z \simeq 6.5$ through Monte Carlo simulations, and we present pilot observations to test the performance of tunable filters to find LAEs and other emission-line candidates. 
    % Methodology | Materials
    We compare the samples obtained using a narrow ideal-rectangular-filter, the Subaru NB921 narrow-band filter, and sweeping across a wavelength range using the ultra-narrow-band tunable filters of the instrument OSIRIS, installed at the 10.4\,m Gran Telescopio Canarias.  %
    We use this instrument for extracting LAE candidates from a small set of real observations. %
    Broad-band data from the Subaru, HST and Spitzer databases were used for fiting SEDs to calculate photometric redshifts and to identify interlopers.
    % Results | Achievements/Contribution | Implications
    Narrow-band surveys are very efficient to find LAEs in large sky areas, but the samples obtained are not evenly distributed in redshift along the filter bandpass, and the number of LAEs with equivalent widths $< 60$\,{\AA} can be underestimated. %
    These biased results do not appear in samples obtained using ultra-narrow-band tunable filters. %
    However, the field size of tunable filters is restricted because of the variation of the effective wavelength across the image.
    % Applications
    Thus narrow-band and ultra-narrow-band surveys are complementary strategies to investigate high-redshift LAEs. %
\end{abstract}
    % Limitatins | Future work


\section{Introduction} \label{sec:intro}

% Define Lyman Alpha Emitters (LAEs) and Lyman Break Galaxies (BLGs).

    % Lyman-break galaxies (LBGs) and Lyman alpha emitters (LAEs) are star-forming galaxies at high redshift, and they are thought to be progenitors of modern galaxies.
    The odd spectral energy distribution of \la Emitter Galaxies (LAEs) convolved with the instrumental profile can introduce unsought biases on the characteristics of the candidate samples extracted from photometric surveys. Particularly, properties such as the the Equivalent Width (EW) and the redshift distribution of the subsequently spectroscopically confirmed LAEs, may be easily affected by the photometric survey instrumentation and methodology used. On the other hand, these surveys yield reliable photometric redshifts for LAEs and Lyman Break Galaxies (LBGs), and thus they are useful cosmological tools, as these objects trace dark matter halos and subsequently the evolution of matter distribution in the universe. Furthermore, LAEs at $z \geq 6.5$ are also important to study the last stages of the reionization epoch \citep{malhotra:2004, kashikawa:2006, shibuya:2012}. Therefore, it is important to develop unbiased alternative strategies to find new LAE candidate samples.

% Narrow band filters
    LBG candidates are selected using the drop-out technique, which consists in comparing images of the galaxy obtained in several broad-band filters that cover contiguous wavelength ranges to both sides of the Lyman break at 912\,{\AA} \citep{steidel:1996}. In a similar way, LAE candidates are usually selected by an excess in a narrow-band filter compared to a nearby broad-band image \citep{cowie:1998}. The later technique is routinely used to find LAE candidates in the Subaru Deep Field \citep{taniguchi:2005, kashikawa:2006, kashikawa:2011} and the Subaru/XMM-Newton Deep Survey Field \citep{ota:2010, ouchi:2010, shibuya:2012}. Narrow-band filters have bandwidths of Full Width at Half Maximum FWHM~$\sim$~100\,{\AA}, and to avoid atmospheric OH emission-lines, narrow-band surveys are confined to a limited number of redshift ranges. An important result of the narrow-band surveys is that the luminosity function of LAEs remains constant between redshifts $3.0 \leq z \leq 5.7$, but evolves dramatically between $5.7 \leq z \leq 6.5$, and maybe beyond \citep{pentericci:2011,hibon:2012,ota:2012a}.

% Ultra narrow band filters
    Despite the utility of narrow-band filters to find LAE candidates, some groups have used ultra-narrow-band filters to perform this task. Thus, \citet{tilvi:2010} and \citet{krug:2012} have employed a filter with a FWHM~$\sim$~9\,{\AA}, installed on the 4\,m Mayall Telescope, seeking LAEs at redshift $z = 7.7$; they expected to find one or two LAEs, respectively, but instead found four LAE candidates, and argued about a possible lack of evolution of the LAE LF for redshifts from 3.1 to 8. Recently, \citet{swinbank:2012} searched for LAEs around two quasars at $z \sim 2.2$ and one quasar at $z \sim 4.5$ with the Taurus Tunable Filter instrument installed on the 3.9\,m Anglo-Australian Telescope, using a bandpass of FWHM~$=$~10\,{\AA}, and found a local number density an order of magnitude higher than what might be expected in the field.

% Characteristics of LAEs
    The spectra of LAEs are characterized by an asymmetric \la line profile with a steep blue cutoff due to absorption by neutral hydrogen, as shown in Figure 3 in \citet{hu:2006}.  At redshifts $z > 6$, hydrogen completely suppresses the continuum of the galaxy at the blue side of the line (Gunn-Peterson trough), while there is a dim continuum at the red side.  Narrow band selected LAEs usually have high line luminosities ($L_{\alpha} > 1.5 \times 10^{42} \ergs$) and large line equivalent widths at rest \citep[$EW_{0} \geq 20$\,{\AA}, e.g. ][]{dayal:2012, mallery:2012}, although the actual limits may substantially vary because of technical constraints and selection strategies \citep[cf. ][]{hayes:2010, ouchi:2008, ouchi:2010}.  These properties provide clues which suggest that LAEs represent an early stage of a starburst in an interstellar medium of very low metallicity and almost free of dust, with a radius of about 2\,kpc and star formation rates around 6\,M\sun\,yr$^{-1}$ \citep{taniguchi:2005}.

% Scattering by neutral hydrogen
    The fraction of \la photons that escape from a high-redshift star forming region is still an active and open topic, as well as an important parameter to account for the observable properties of LAEs. Neutral hydrogen resonantly scatters the \la photons, changes their escape paths and hinders the realization of the photons that are absorbed by dust. \citet{ono:2010} place an upper limit of 20\% escaping photons at $z = 6.6$, but at lower redshifts \citet{blanc:2011} and \citet{ciardullo:2012} find that the photon escape fraction may be as high as 100\%. \citet{hayes:2010} consider the \la luminosity density at $z = 2$ significantly underestimated from its intrinsic value, due to the fact that only 1 in 20 photons, on average, reach the telescope; this will specially happen at $z > 6$, where the neutral fraction of the intergalactic medium may cause significant suppression of the \la line \citep{santos:2004, hayes:2006, dijkstra:2007}. %
% Clumpy dust as enhancer
    On the other hand, \la photons may be less attenuated than non-resonant radiation under suitable conditions within a multiphase scattering medium \citep{neufeld:1991, hansen:2006, finkelstein:2008}. %
    %
    % This line resonance may introduce a bimodality on the age of observed LAEs \citep{finkelstein:2011}, with objects that are either very young ($<15\,\rm{Myr}$) or old ($> 450\,\rm{Myr}$). %
    For \citet{finkelstein:2011}, dust geometry shapes the spectral energy distributions, and has a major influence in the observed EW at the rest frame of the \la line ($EW(\la)$) distribution in high-redshift LAEs. %
% Kinematics
    Finally, the presence of outflows can increase the fraction of \la photons that escape from the star forming region \citep{kunth:1998, tapken:2007, verhamme:2006, verhamme:2008, atek:2008, atek:2009}.

% LAEs-LBGs connection
    If the observational aspects of LAEs depend on geometrical effects (dust spatial distribution), orientation, and the presence of outflows, the separation between LBGs and LAEs becomes somewhat diffuse. Thus, \citet{kashikawa:2007} classify galaxies with contrasts as low as $EW(\la) > 10$\,{\AA} as LAEs, while recently \citet{krug:2012} consider even lower limits ($EW(\la) \gtrsim 4.8$\,{\AA}) for $z \sim 7.7$ LAEs. Besides, other observations indicate that the fraction of low ultraviolet luminosity LBGs ($M_{\rm{UV}} > -20.5$) with $EW(\la)>50$\,{\AA} is about one half \citep{stark:2010, stark:2011, pentericci:2011, schenker:2012, vanzella:2011, ono:2012}. The presence of \la lines in some LBGs provides a strong evidence of the connection between these objects and LAEs. Eventually, this connection will promote an effort to build theoretical models to explain the underlying physics of LAEs and LBGs \citep[e.g ][]{shapley:2001, dayal:2012, forero-romero:2012}.

% Previous surveys
    % Nilsson, K. K., A. Orsi, C. G. Lacey, C. M. Baugh & E. Thommes 2007, A\&A, 474, 385
    % One of the most promising ways of detecting very high-redshift (z & 5), star-forming galaxies is via narrow-band imaging surveys targeting Lyman-a (Lya). In particular, redshifts z ~ 5.7 and 6.5 have been extensively surveyed by several groups (e.g. Ajiki et al. 2003; Hu et al. 2004; Shimasaku et al. 2005; Ouchi et al. 2005, 2007; Malhotra et al. 2005; Taniguchi et al. 2005; Tapken et al. 2006; Kashikawa et al. 2006). The current redshift record for a spectroscopically con?rmed Lya emitter (LEGO Lya Emitting Galaxy-building Object; see M{\o}ller & Fynbo 2001) is z = 6.96 (Iye et al. 2006) although Stark et al. (2007) have suggested the discovery of two LEGOs at z = 8.99 and 9.32. The reason why narrow-band surveys are restricted to a discrete number of narrow redshift windows is the night sky OH emission-lines

%% PROGRAM

    We are conducting a blind search on selected gravitationally lensing galaxy cluster sky fields to obtain samples of LAEs not subjected to possible biases imposed by the methodology of narrow-band filters. Hence, we are obtaining ultra-narrow-band images using tunable filters of the instrument OSIRIS (\otf) attached to the 10.4\,m Gran Telescopio Canarias (GTC). %
    \otf is optimized for line flux determination and thus it can be called a Star Formation Machine \citep{cepa:2003}. This characteristic makes \otf a powerful instrument to detect faint young galaxies with accurate photometric redshift estimates. %
    Therefore, apart from LAEs, we also expect to obtain other high-redshift candidates, such as LBGs, AGNs and line emitters.

%% Paper organization

    This paper is  organized as follows. %
    Section~\ref{sec:bias} introduces the bias produced by the filter bandwidth on the photometric selection of LAEs. %
    Section~\ref{sec:sample} describes the generation of Monte Carlo data for simulated LAE spectra. %
    Section~\ref{sec:nband} presents the LAE recovered samples obtained using narrow-band filters on the simulated data, exemplified by an ideal rectangular filter and the NB921 Subaru filter. %
    Section~\ref{sec:tf} describes the methodology used to search for LAE candidates using \otf, and presents the sample recovered from the simulated data. Our methodology is then applied on a pilot observation of real data obtained at the GTC to select preliminary candidates at redshifts $z \simeq 6.5$. %
    In Section~\ref{sec:archive} we cross-check our \otf preliminary candidates with photometric archive data and models of galaxies at high redshift to study the Spectral Energy Distribution (SED) and to improve classification.
    Section~\ref{sec:discuss} presents the discussion of the results obtained with simulated and real data. Finally, %
    Section~\ref{sec:conclus} presents our conclusions.

%% Cosmology

    Throughout this paper we assume a cosmological model with $H_0=70\,\mathrm{km \, s^{-1} \, Mpc^{-1}}$ ($h=0.7$), $\Omega_{\rm{M}} = 0.27$, and $\Omega_{\Lambda} = 0.73$.

\section{The asymmetric continuum bias} \label{sec:bias}

    The spectrum of a LAE at $z \sim 6.5$ shows a sharp decay due to the absorption of \la photons by neutral hydrogen that truncates the blue part of the emission line. From the observational point of view the resulting line is asymmetric, and the signal is circumscribed to wavelengths longer than \la.

    The detection of dim objects is a difficult task and very often the results may vary depending on subtle details. In the case of LAEs, the spectral asymmetry of the host galaxy continuum background with respect to the \la line may enhance the detection of objects with the largest contribution of the continuum to the total signal. If we use a narrow-band filter with a FWHM comparable to the EW of the \la line, the contribution of the continuum to the total recorded intensity may change significantly depending on the redshift of the object. Figure~\ref{fig:bias} shows an example of this effect on an ideal rectangular filter of a width of 132\,{\AA}, similar to the FWHM of the NB921 Subaru filter. The figure shows two simplified LAE spectra at slightly different redshifts (6.51 and 6.61). The objects have an observed FWHM of 10\,{\AA} for the \la line, and an identical continuum level. For the LAE at the lowest redshift, the \la line lies on the short-wavelength edge of the filter, and its EW is 2.5 times the wavelength range of the filter (i.e., 330\,{\AA} in the observer's frame or  44\,{\AA} in the rest-frame), which corresponds to a difference of one magnitude between the line and the continuum, a criterion often used to select LAE candidates in narrow-band surveys \citep[e.g. ][]{ouchi:2010, kashikawa:2011}. For the LAE at the highest redshift, to reach the same signal through the rectangular filter, the intensity of the \la line (or its EW) should be approximately 34\% higher than the line intensity of the former object.

\input{fig2}

    The bias introduced by the asymmetric continuum profile at both sides of the \la line is more pronounced as the EW decreases. The observed-frame EW for LAEs have values $\sim 100$\,{\AA}, similar to a narrow-band filter FWHM. Therefore, the asymmetric continuum may affect the shape of the LF of LAEs inferred from narrow-band surveys. As objects are less luminous, the volume actually studied depends on the amount of continuum in the range of the filter.

\section{Sample simulation} \label{sec:sample}
%% In silico

%% SPECTRUM PROFILES

% Background
    % *******************
    % Referee 1a: Shimasaku et al. 2006 -- Ouchi et al. 2008
    % *******************
    \citet{shimasaku:2006} and \citet{ouchi:2008} have analyzed the redshift distribution of LAEs in Subaru's narrow-band filters using Monte Carlo mock samples of LAEs.
    % *******************
% Our simulations
    For the purpose of comparing the characteristics of samples obtained with the \otf and other instruments, we have also built a simulated sample of LAEs. This sample was made by modeling the LAE spectra with a superposition of an asymmetric triangular profile for the \la line plus a continuum, which is a simplified version of the profile model by \citet{hu:2004}. The \la line profile consists of a sawtooth with the steeper inclination at the line's blue side. At wavelengths shorter than the \la line, the continuum is zero, and it has some constant non-zero value at wavelengths equal or larger than the \la line. Altogether, we have simulated the spectra of \nsimul LAEs.

% LF
    We have used the Schechter function to model the LF of LAEs at $z \simeq 6.5$:
    \begin{equation}
    %\begin{equation}\label{eq:schechter}
        \phi  = {\phi _*} \left( {\frac{L}{{{L_*}}}} \right)^\alpha \exp{\left( {\frac{-L}{{{L_*}}}} \right)} \, \frac{{dL}}{{{L_*}}} ,
    \end{equation}
    with the parameters given by \citet{kashikawa:2011}: $\alpha=-1.5$, $\log(L_*/h^{-2}\,\mathrm{erg\,s^{-1}})=42.76$ and $\log(\phi_*/h^3\,\mathrm{Mpc^{-3}}) = -3.28$.

    The simulated LF sample was computed through the inversion method of the cumulative Schechter function, integrating between \lolim\ and \uplim\ \ergs, a luminosity range that broadly includes the observed LAEs at redshifts $z \approx 6.5$ \citep[e.g. ][]{kashikawa:2011}.

% FWHM
    A random FWHM sample for the \la line has been computed according to the distribution parameters inferred from Table~2 in \citet{kashikawa:2011}. These authors have measured the observed FWHM of 28 $z\approx 6.5$ LAEs with values between 5.04\,{\AA} and 25.2\,{\AA}, with mean of 13\,{\AA} (428\,km\,s$^{-1}$) and a standard deviation of 5\,{\AA}. The FWHM sample has been shifted to the rest-frame of the LAEs to build the profile of the emitted \la line.
		
% EW
    The random EW sample at rest-frame ($EW_0$) has also been computed like the luminosity, using the inversion method on the cumulative function extracted from Figure~11 in \citet{kashikawa:2011}. %, and fitted using a lognormal cumulative function \citep[e.g.][]{gaskell:2004, collet:2003}.
    % *******************
    % Referee 2: Blanton and Lin 2000 -- Shapley et al 2003
    % *******************
We tried to fit the EW distribution using exponential cumulative functions, but we obtained a better fit using the lognormal cumulative function. %
Lognormal EW distributions also fit the [\ion{O}{2}]$_{\lambda3727}$ equivalent widths in the local ($z<0.2$) universe \citep{blanton:2000}, and has been employed by \citet{shimasaku:2006} to characterize the EW of LAEs at redshift  ($z\simeq5.7$). However, the \la EW distribution in LBG at $z\sim3$ does not show lognormal profile \citep{shapley:2003}, possibly because in LBGs \la is observed both in emission and in absorption, depending on the object. In any case, accurate EW measurements are difficult mainly because of the low level of the galaxy continuum, and thus the actual shape of the \la EW distribution is still an open subject.
    % *******************
    % We have used a log-normal function to fit the data and also the
    % inversion method on the cumulative function.
    % A variable might be modeled as log-normal if it can be thought
    % of as the multiplicative product of many independent random
    % variables each of which is positive.
    % Ref. Wikipedia
    % http://en.wikipedia.org/wiki/Log-normal_distribution
    % This $EW_0$ has a mean value of 123\,{\AA}
    The $EW_0$ in our simulations has a median value of 68\,{\AA}, slightly below the value of 74\,{\AA} reported by \citet{kashikawa:2011}. By construction, the lower values of the simulation were restricted to $EW > 0$. Extreme values are useful to test the instrumental response and to set constraints to the distribution of real objects. If the simulations show that the instrument reaches a certain parameter or combination of parameter values, but the corresponding objects have not been observed, it is evidence that there are few if any of them. On the other hand, if the simulations show that the instrument cannot detect objects with certain combinations of parameter values, it remains an open question whether these objects exist or not. In any case, because the simulation is based on empirical parameter distributions, extreme values are very improbable, and they cannot significantly alter the statistical results of this study. For example, in our simulation we find 69 objects with $EW_0 < 8$\,{\AA}, and 77 with $EW_0 > 600$\,{\AA} from 2708 \otf recovered LAEs; of course these numbers depend on the EW distribution adopted, which may be critical for the low EW regime.

    To set thresholds for the EWs of LAEs is not a simple task. In some cases the continuum may be undetected, and the maximum threshold cannot be determined. In the case of the minimum threshold, \citet{stark:2010} have proposed that the equivalent width at the rest frame should be $EW_0 > 55$\,{\AA} for LAEs. %
    %
        % FROM DAYAL \& Ferrara (2011): ...using the Lyman Alpha equivalent width selection threshold EW > 55{\AA} of Stark et al. (2010), the LAE fraction increases towards the faintest LBGs. However, for the canonical value of EW > 20{\AA} , all LBGs with -23 < M_{UV} < -19 would be identified as LAEs at $z \sim 6$; the fraction of LBGs showing a Lya line decreases with redshift due to the combined effects of dust and reionization.
    %
    More recently, these authors have proposed a value of 25\,{\AA} \citep{stark:2011} that approaches the widely accepted minimum threshold of $EW_0 > 20$\,{\AA}. %
    %
        % FROM STARK ET AL (2011): The rest-frame EWs for the i'-drops range between 9.4 {\AA} and 350 {\AA}. The vast majority of the emission lines are detected with high significance. Even so, we take a conservative cut, limiting our analysis to those sources with rest-frame EWs greater than 25 {\AA} and SNR > 7
    %
    The intensity of the \la line and its EW may depend on intrinsic properties of the LAEs or on geometrical attributes (see {\S}\,\ref{sec:intro}). %
    Moreover, the adopted threshold is a figure that probably depends also on the observational limits adopted to separate candidates efficiently. Thus, \citet{stark:2011} note that the measured rest-frame EWs of the \la line for 13 spectroscopically confirmed LAEs range between 9.4\,{\AA} and 350\,{\AA}. Therefore, we have conserved even the most extreme values for the EW in our simulations. Besides, these values may be useful to make extreme-case differences between the respective methodologies to detect LAE candidates.

% CONTINUUM
    For each simulated LAE, once the power of the \la line ($L_{\la}$) and its $EW_0$ have been assigned, it is possible to calculate the spectral power $\Phi_c$ of the continuum:
    \begin{equation}\label{eq:specpow}
        \Phi_c = \frac{L_{\la}}{EW_0} .
    \end{equation}
    This spectral power is used to compute the spectral shape of the continuum at wavelengths $\lambda \geq \lambda_{\la}$. For wavelengths $\lambda \leq \lambda_{\la}$, the continuum is assumed to be zero, as the optical depth for photons with wavelengths shorter than the \la line is large. Equation~\eqref{eq:specpow} is also used to build a continuum of reference for all the wavelengths of interest in our analysis, which will be used later to estimate a broad-band continuum emission necessary for the criterion of detection.

% FLUXES
    The observed spectra for both, the \la line and the continuum, have been calculated from the corresponding spectral luminosities taking into account the redshift for each simulated object. These lines and continuum spectra have been added in order to obtain the observed flux, from which we will get the photometry later in our analysis. We have constructed another spectrum of constant intensity for each object, calculated from the corresponding continuum of reference.

\input{fig3}

\section{Narrow-band filters} \label{sec:nband}

    In this section we analyze the outcome produced by narrow-band filters applied to our simulated sample. %
    % *******************
    % Referee 1b: Gronwall et al 2007 
    % *******************
    \citet{gronwall:2007} have noted the effect of a nonsquare bandpass in the definition of survey volume and sample's flux calibration for large photometric selected samples of emission-line galaxies. For these authors, the volume of space sampled is a strong function of line strength: objects with bright line emission can be detected even if the \la line lies in the wings of the filter, but for weak \la sources the line must lie near the bandpass center to be noticeable. Furthermore, ignoring the actual position of the line in the bandpass prevents a precise estimate of the effective filter transmission in the line position, affecting the flux calibration. 
    % *******************
    %
    Being aware of these constraints, we consider two narrow-band filter profiles. First, an ideal rectangular filter to study the effect of a general pass-band filter on the detection of LAEs, and secondly a filter with a response similar to that of %
        % Suprime-Cam has ten fully-depleted-type 2048x4096 CCDs manufactured
        % by Hamamatsu Photonics KK (S10892-01), which are arranged in a 5x2
        % pattern, providing a field of view of 34' x 27'.
    Subaru NB921 attached to the Suprime-Cam, to study the effect of a real-life filter profile.

    Figure~\ref{fig:histograms} shows the distribution of the simulated sample, the ideal and the NB921 Subaru filters, and the \otf (see \S\ref{sec:tf}). The simulations yield a number of approximately 400 objects per wavelength bin (each bin with a width of 25\,{\AA}).

\subsection{Ideal rectangular filter} \label{sec:ideal}

% Filter width
    The ideal rectangular filter is defined such as:
    \begin{equation}\label{eq:ideal}
        % MATLAB variables:
        %   eNB921 = 9196
        %   fwNB921 = 132
        %   eNB921 +/- fwNB921/2 = 9130, 9262
        T_{\lambda} =
          \begin{cases}
            1 & \text{if } 9130 \leq \lambda \leq 9262\, \text{\AA}, \\
            0 & \text{otherwise,}
          \end{cases}
    \end{equation}
    where $T_{\lambda}$ is the transmission of the filter as a function of the wavelength $\lambda$. The bandwidth of the filter defined above is 132\,{\AA}, which equates the FWHM of the Subaru NB921 filter (see {\S}\,\ref{sec:subaru}).

% Detections
    Simulated LAEs are recovered according to two criteria, one for source detection and the other for LAE candidate discrimination. The detection criterion is fulfilled when the irradiance in the narrow-band filter is $\geq 5 \times 10^{-18} \, \ergscm$, which corresponds to the Subaru flux limit \citep{ouchi:2010}. The discriminant criterion is a contrast condition between the narrow-band and the broad-band fluxes. This criterion is in fact a simplified version of the \citet{ouchi:2010} color criterion, but is sufficient to analyze the simulated data. In our case, the condition that must be accomplished consists of the narrow-band flux being at least a magnitude brighter than the (adjacent) broad-band flux, which provides a continuum of reference. Computationally, this criterion leads to the flux of the object measured through the filter being at least 2.5 times larger than the flux measured from the constant continuum of reference computed in {\S}\,\ref{sec:sample}.

% Results
    Results are shown in Figure~\ref{fig:histograms}b. Roughly, due to the asymmetric continuum bias, a larger number of detections at shorter wavelengths are expected, and decays smoothly at longer wavelengths. Small departures from this behavior are due to random deviations in the original simulated sample redshift distribution. Thus, most detections are scattered over the filter pass-band, but some simulated objects with their \la peaks lying at shorter wavelengths are recovered.

    Only around a fourth part of the wavelength range in the leftmost bin in Figure~\ref{fig:histograms}b lies inside the filter profile, and almost half of the recovered objects in this bin correspond to simulations with the \la line peak located inside the filter pass-band. In fact, most recovered objects in this bin (83 out of 160) have the peak of the \la line at wavelengths shorter than the filter window, but with a fraction of the long wavelength queue of this line inside the filter. Note that the detection of these objects will depend on the line parameters (peak position, FWHM and intensity), and that the recorded signal will be diminished by the loss of the line flux outside the filter window, and thus these objects tend to have larger EWs that make detections easier. On the other hand, the smooth decay of the number of recovered LAEs along the filter pass-band is a result of the smaller quantities of the continuum emission lying inside the filter as the \la line peaks at longer wavelengths. For objects with their \la line near the long wavelength extreme of the filter, the contribution of the continuum to the total flux in the band is negligible, and part of the long wavelength queue of the line may also lie outside the filter window, yielding a steep drop in the number of detections. Therefore these objects also tend to have large EWs to compensate for their total loss of flux.

\subsection{Subaru filter} \label{sec:subaru}

% CCD and Transmission
% ** This has been deleted as the CCD is no further taken into account **
%    The transmission of the Subaru NB921 filter has been combined with the response of the Hamamatsu detector to account for the instrumental response. The spectral response of the detector corresponds to frequencies much lower than the filter, and thus its effect on the total FWHM is negligible.

% FWHM
    NB921 filter is characterized by an almost Gaussian profile with a central wavelength at 9196\,{\AA} and a FWHM of 132\,{\AA} \citep{kashikawa:2011}, that corresponds to a spectral resolution of 70. The transmission profile is available from Subaru.\!\footnote{\anchor{http://www.naoj.org/Observing/Instruments/SCam/sensitivity.html}
    {http://www.naoj.org/Observing/Instruments/SCam/ \\ sensitivity.html}} A close inspection of this profile shows that the maximum of the recovered LAEs is slightly offset (9183\,{\AA}) with respect to the filter peak, in agreement with the \la wavelength distribution of confirmed LAEs reported by \citet{kashikawa:2011}.

% Detections & Results
    We used the same detection criteria as in the case of the ideal rectangular filter. The results are shown in Figure~\ref{fig:histograms}c. The number of recovered LAEs is restrained by the filter profile along with the same effects that have been noted in the ideal rectangular filter. On the one hand there is the decay of the continuum contribution for the objects with the peak of the \la line closer to the long wavelength limit of the filter, on the other hand there also is the loss of the long wavelength queue of the \la line, which spreads to the low transmission region of the filter profile. Therefore, the long wavelength queue of the line observed through the Subaru filter is damped by the bell-shaped transmission, in agreement with \citet{shimasaku:2006} and \citet{ouchi:2008}.

\section{Tunable filters} \label{sec:tf}

Throughout this work, a distinction is drawn between a \emph{frame}, corresponding to one set of data read from the CCDs; an \emph{image}, a number of frames at the same etalon settings which have been combined for analysis; a \emph{field}, a stack of images of the same area of sky at different etalon settings.

\subsection{OSIRIS tunable filters} \label{sec:otf}

%% Transmission
    Basically, the imager/spectrograph \otf is a low resolution Fabry-P\'{e}rot spectrograph which consists of a blue and a red arm. Our observations were performed with the red arm, that can be centered at any wavelength between 6500 and 9300\,{\AA}, and we observed using a fixed bandpass of 12\,{\AA} which corresponds to a spectral resolution of about 770, %
    % for wavelengths larger than 8500\,{\AA}, %
    % Imposed by the design of the order-sorting filters (USER-Manual page 34)
    a set of order-sorter filters to avoid contamination by neighboring orders, and a detector array consisting of %
    two MAT 4k$\times$2k CCDs \citep{cepa:2003}. % (~9.2 arcsec gap 2) from same Si wafer
    The 12\,\AA\,\,bandpass is the only one currently available for wavelengths around 9200\,\AA.

%% Observing strategy
    The observing strategy consists on sweeping a selected spectral range using steps of a half of the FWHM (in our case, 24 images shifted 6\,{\AA} between 9122 and 9260\,{\AA}) to avoid aliasing. To apply this methodology to our simulated data, we have modeled a set of 24, approximately Lorentzian shaped, tunable filters using the following approximation that relates wavelengths and transmissions of the \otf \citep[][ eq. 3.14]{cepa:2011}:
    \begin{equation}\label{eq:osiris}
        T \approx \Bigg\{1+\left[
        \frac{2(\lambda-\lambda_0)}{\delta\lambda}
        \right]^2\!\Bigg\}^{-1} \!\!\!\!\!\! ,
    \end{equation}
    where $\delta \lambda$ is the filter FWHM and $\lambda_0$ is the central wavelength.

    % LAE (a) at z=6.6 is in fact at 6.5609.
    Figure~\ref{fig:osiris} shows a simulated LAE at redshift $z=6.6$ (a), the set of \otf filtered spectra for this simulation (b), and the photometric output (c). This particular simulation has a rather small $EW_0 = 17.8$\,{\AA}, which is adequate to make the continuum more apparent in the plot. In panel (c), the short wavelength edge of the line aliases because the spectrum varies at a higher frequency than the \otf spectral resolution. Nonetheless, the asymmetry of the line profile is still noticeable.

\input{fig4}

%% Detection criterion

    In this paper we have adopted two criteria to retrieve LAEs from the simulations, based on the characteristics of \otf. The first is a detection condition that imposes an irradiance \mbox{$\gtrsim 4 \times 10^{-18} \, \ergscm$} for the field narrow-band image. This field is built through the sum of the 24 ultra-narrow-band filters to provide a wider synthetic filter \citep[band synthesis technique, ][]{cepa:2009}. The second condition consists of a ratio between the maximum and minimum of the 24 filters larger than 2.5 (i.e., one magnitude). These conditions are equivalent to those imposed to the Subaru survey.

\input{tab1}

%% Results
    Results are shown in Figure~\ref{fig:histograms}d. The distribution of the simulated LAEs recovered using \otf expands the full range of wavelengths between 9122 and 9260\,{\AA}. A visual inspection of Figures~\ref{fig:histograms}a and \ref{fig:histograms}d shows that the shape of the distribution in this range of wavelengths closely resembles that of the simulated LAEs, without any apparent bias introduced by the filter profiles. Also notice that almost all the simulated objects in this range are recovered. In addition, there also are some retrieved LAEs with their \la line peak ($\lambda_{\la}$) outside the 9122 --- 9260\,{\AA} range, which is a consequence of the \otf Lorentzian profile. These objects tend to have large EWs; and those with $\lambda_{\la} > 9260$\,{\AA} tend to be brighter, too.

\subsection{\otf pilot observations}\label{sec:observations}

% OSIRIS-TF Log

    Photometry was carried out at the GTC using \otf %
    % (Cepa, 1998, 2000;Cepa et al., 2003)
    with FWHM of 12 {\AA} at five contiguous wavelengths separated by steps of 6\,{\AA}, and a field of view free of adjacent orders of about 8 arcminutes on each side. The observations were performed with central wavelengths $\lambda_c = \text{9122\,{\AA}, 9128\,{\AA}, 9134\,{\AA}, 9140\,{\AA} \& 9146\,{\AA}}$. The observation run was done on \obsdate. A binning of $2 \times 2$ was used in fast (standard) readout mode (200 kHz), with three dithered exposures per wavelength of 210\,s each, and separated by a triangular offset pattern of 10 arcseconds to eliminate diametric ghosts during data reduction, each called a frame. The night was photometric and the seeing varied from 0.75 to 0.82 arcseconds during the observation run.
    
    Following the observation of the cluster in all the wavelengths, a standard star was observed with the same instrumental settings (but different exposure times) for flux calibration.

    Standard IRAF procedures for bias subtraction were used on the data. Super-flats (a flat generated by averaging the sky from scientific images where sources have been masked) were created from and divided to the scientific frames due to the unevenly-lit dome flats. Because of the TFs small bandpass and position-dependent wavelength, all observations contain sky (OH) emission rings which were subtracted to all frames with the IRAF package TFred,\!\footnote{Written by D. H. Jones for the Taurus Tunable Filter, previously installed on the Anglo-Australian Telescope; \url{http://www.aao.gov.au/local/www/jbh/ttf/}} %
    % http://www.aao.gov.au/local/www/jbh/ttf/adv_reduc.html
    which estimates the sky background, including the sky rings which are several arcminutes in diameter. The three frame offsets of each wavelength were aligned and combined to generate an image with a total exposure time of 630\,s, these were later convolved to the worst seeing of 0.82 arcseconds.

\input{fig5}

% Searching candidates with the deep image
    Also, a field was created convolving to the worst seeing of 0.82 arcseconds, and combining all aligned images of different wavelengths (band synthesis technique). This field has a total integration time of 3150\,s, and yields a detection limit irradiance of $9 \times 10^{-18} \, \ergscm$ integrated over the full wavelength range (36\,\AA) of the synthetic band. We used SExtractor \citep{bertin:1996} to make a catalog of detected sources in the field. We excluded sources from the catalog that were too bright for high-redshift galaxies. Photometry on each monochromatic image was performed using an aperture of 1.5\,arcseconds %
    % originally  2.97\,arcseconds (11.7\,pixels)
    using the positions gathered by SExtractor. Then we selected possible candidates based on the maximum vs minimum flux ratio, excluding all those sources with a ratio below 2.5 (i.e., one magnitude). The remaining objects were carefully inspected by eye to reject faint cosmic rays, ghost residuals, and source contamination by nearby companions or located too close to the edge of the image; the region around the gap between the detectors was particularly clumped with fake detections. Finally, we selected those candidates that showed a photometric profile similar to those expected for LAEs and LBGs with either the \la line or the Lyman Break lying, at least partially, inside the observed wavelength range. Given the likely range of Ly$\alpha$ emission-line widths, and the wavelength sampling, we expect to observe this line in more than three adjacent passbands.

    Table~\ref{tab:candidates} summarizes the data for the candidates. Column~1 identifies the \otf candidate. Columns~2 and~3 show the right ascension and declination coordinates, respectively. Column~4 shows the redshift obtained from the peak of the alleged \la line or Lyman break. Column~5 is the total irradiance corresponding to the field. Column~6 and 7 present the classification of candidates obtained using the \otf data, and the SED fitting to Subaru, HST and Spitzer photometrical data. 

    Figure~\ref{fig:observations} shows the \otf images for each LAE and LBG candidate. Each row in this figure corresponds to a different object located at the center of a guiding circle. Despite candidate \dlla being near the upper border of the clipped image, it does not affect our analysis. This is also the case for the candidate \lbga, which is near the diffuse border defined by the dithered gap between the detectors. The first column of frames in Figure~\ref{fig:observations} shows the field exposure obtained by piling up all the individual images; the rest of columns show the candidates observed at the wavelength identified in the heading row.

\input{fig6}
        
\input{fig7}

    Figure~\ref{fig:LaeCandidate} shows three LAE candidates observed with the \otf at the GTC, and selected simulations extracted from our database that resembles the real data, rather than a fitted model for each object. The shift in wavelengths between the observations and the model is due to the dependency of the tuned wavelength of the filter on the distance $r$ to the optical center in the \otf images \citep[ in preparation]{gonzalez:2013}:\footnote{\url{http://gtc-osiris.blogspot.com.es/}}
    % \citep{gonzalez:2013}:
    %
    \begin{equation}\label{eq:wavelengthshift}
        % Equation in \citet{cepa:2011}
        % \Delta \lambda = -7.9520 \times 10^{-5} \lambda_c r^2,
        % New equation (no reference yet)
        \lambda  = {\lambda _c} - 5.04{r^2} \!\!\! , % + {a_3}(\lambda){r^3},
    \end{equation}
    where $\lambda_c$ is the wavelength at the optical center expressed in {\AA}, and $r$ in arc\-minutes.
    %, and $\lambda$ and the $a_3(\lambda_c)$ term should be calculated iteratively:
    %    \[
    %    {a_3}(\lambda) = 6.0396 - 1.5698 \cdot {10^{ - 3}}\lambda  + 1.0024 \cdot {10^{ - 7}}{\lambda ^2}.
    %    \]

    The redshifts for these LAE candidates were calculated identifying the brightest photometric point with the position of the \la line. For the peak-missed LAE candidate shown in  Figure~\ref{fig:LaeCandidate}c, the redshift estimate is therefore an upper limit (see Table~\ref{tab:candidates}). In fact the fraction of candidates with either upper or lower redshift limits are expected to be relatively high. This subject, and the possible contamination of the candidate sample by foreground galaxies, will be addressed in \S\ref{sec:archive} and \S\ref{sec:discuss}.

    The two objects shown in Figure~\ref{fig:DubCandidate} have a photometric profile compatible with either LAE or LBG candidates. Their respective redshift lower limits are shown in Table~\ref{tab:candidates}. The photometric profile for the object shown in Figure~\ref{fig:DubCandidate}a may be contaminated by residuals of the sky-line subtractions at wavelengths below 9050\AA. As in the previous figure, simulated objects extracted from our database are also plotted for comparison purposes.

    Objects shown in Figure~\ref{fig:LbgCandidate} are two LBG candidates. Both objects show a steep increase in flux that is held at longer wavelengths, as it is expected for the continuum emission of LGBs.

\section{Subaru, HST and Spitzer archive data}\label{sec:archive}

% Discussion about HST and Spitzer data

%% PUBLIC DATABASES

    We have searched for additional photometric data for our LAE and LBG candidates in astronomical public databases, and have found some deep images in the Subaru, Hubble Space Telescope (HST) and Spitzer archives that cover, at least partially, the \otf field around \cluster. %
    We used these data along with our observations to fit the SED of the \otf candidates, obtaining a more accurate classification (see Table~\ref{tab:candidates}). %
    Table~\ref{tab:fluxes} shows the fluxes in the different bands of Subaru/Suprime--Cam, HST/WPFC2 and Spitzer/IRAC data. This table also includes the fluxes of the \otf band synthesis images, that correspond to a filter of $\rm{FWHM}= 36$\,{\AA}. In the case of \otf fields and HST fluxes, we found discrepancies based on Subaru calibrations.  We then used stars in the field to recalibrate the \otf and HST data to the Subaru photometry.
    
    \subsection{Subaru observations}
    
\input{fig8}
        
    The Subaru/Suprime--Cam data were obtained from the Subaru-Mitaka Okayama-Kiso Archive System (SMOKA).  They consist of data in filters $V$, $i$\textquotesingle\ and $z$\textquotesingle\ observed in 2009.  The wavelength range of the $z$\textquotesingle\ filter includes our \otf observations, resulting in a helpful band to estimate the continuum.  Data was reduced following the Suprime--Cam Data Reduction software (SDFRED2). %\footnote{A \anchor{http://www.naoj.org/Observing/Instruments/SCam/sdfred/v2.0/sdfred2_2p0e.pdf}{http://www.naoj.org/Observing/Instruments/SCam/sdfred/v2.0/sdfred2\_2p0e.pdf}} 
    The images were flat fielded, matched in PSF size for a predetermined target FWHM, scaled and combined. The total integration time for $V$, $i$\textquotesingle\ and $z$\textquotesingle\ filters was 5040, 1920 and 1800 seconds, respectively. The photometry for the candidates was performed with the \emph{qphot} IRAF package, fluxes were derived by measuring inside a 1.5 arcsecond aperture.

\otf candidates \laea, \laeb, \laec and \dlla are seen in the $V$, $i$\textquotesingle\ and $z$\textquotesingle\ filters, pointing at possible interlopers (i.e., a galaxy at a lower redshift and with spectral features resembling those of LAEs over a limited range of wavelengths) or LBG candidates.  Candidates \dllb, \lbga and \lbgb are not seen in at least one Subaru/Suprime--cam filter, but their best SED fit suggests they are all candidate interlopers.

    \subsection{HST observations}

    The HST/WFPC2 data were obtained from the Multimission Archive at the Space Telescope Science Institute (MAST). They consist of two programs: ID5991 (filters F702W and F814W), and ID6745 (filters F606W and F814W). The proposals cover slightly different fields. Only two of the seven candidates fall in the images, namely, \laec in the field of the proposal ID5991, and \lbga in the field covered by ID6745.

    %The data reduction process has been presented in \citet{verdugo:2007}.

    The image reduction was performed using the IRAF/STSDASS package. First, a warm-pixel rejection was applied to the images using the IRAF task \emph{warmpix}. The cleaned images were then combined with the task \emph{crrej} to remove cosmic-rays hits. Finally,  the background was subtracted and the WFPC2 chips were combined using the task \emph{wmosaic}. The total integration times for filters F702W and F814W were 2400\,s and 2600\,s, respectively for the proposal ID59991. For the proposal ID6745 the total integration times were 3300\,s and 3200\,s for filters F606W and F814W, respectively. The photometry for the seven candidates was performed in the HST filters with the IRAF package \emph{apphot}. The magnitudes were derived by measuring fluxes inside a fixed circular aperture of 7 pixels ($\sim 0\farcs7$). \otf candidates found in these filters suggest an interloper nature.

    \subsection{Spitzer/IRAC observations}

    A Spitzer/IRAC observation was gathered from the Spitzer Heritage Archive (SHA). The Astronomical Observation Request (AOR) number 18626048 (program Name/ID KTRAN-MS2053/\linebreak[0]30642, P.I. K.-V. Tran) was retrieved for this work. It is a four-channel (3.6, 4.5, 5.8 \& 8\,\micron) IRAC map mode observation consisting of 3 rows and 1 column  with a 42-point cycling dither pattern,  and ``medium'' scale (median separation between dither positions is  53 pixels; the IRAC pixel scale is  approximately the same, $\sim$\,1.2\,arcsec in the four camera bands). The resulting map footprint is a roughly rectangular area of  22.9\,$\times$\,8.6\,arcmin centered at $\alpha$\,$\approx$\,314.113$^{\circ}$, $\delta$\,$\approx$\,$-$4.670$^{\circ}$. The major axis is oriented at P.A.\,$\approx$\,168.2$^{\circ}$.

\input{tab2}

    The data reduction of the four IRAC channels was performed using MOPEX v18.4.9, using as starting point the Corrected Basic Calibrated Data (CBCD). The IRAC CBCD data differs from the standard BCD products in the mitigation of several instrumental artifacts, including stray light, muxstripe, banding, muxbleed (electronic ghosting), column pulldown and jail bars.  The \emph{Overlap} and \emph{Mosaic} pipelines were run to create mosaic images from the individual images. The former removes image background variations due to foreground light sources, while the later removes defects and spurious pixels, reassembles the data onto a common pixel grid, and combines them into a mosaic with a corresponding noise map.

    The four-channel photometry of the sources was performed by means of the Astronomical Point source EXtractor (APEX) tool provided within the MOPEX software. The User List Multiframe mode was used. In the APEX Multiframe mode, required for the IRAC instrument,  the extraction is carried out simultaneously on the stack of input images rather than on a single mosaic image. The position and flux density estimate is provided for each detected source. The User List  mode was chosen, providing APEX with an input list of object positions rather than using the Detect module to automatically find the sources in the field. Point-Response Function (PRF) fitting and aperture photometry was done on the input list objects. The centroids of the photometrical profiles were allowed to move slightly with respect to the input positions, resulting in small shifts between 0\farcs05 and 0\farcs6, except for \dllb (1\farcs3).
    % The results are shown in table \ref{iractab}. For each object, the point source flux and its uncertainty are shown for each IRAC band (when available).
    Only detections with SNR\,$>$\,3 have been considered.
    % The centroid displacements relative to the input source position have been also included.

\subsection{SEDs and photometrical redshifts}

    We have used an updated version of HyperZ \citep{bolzonella:2000} and synthetic spectral templates of galaxies \citep{bruzual:1993, bruzual:2003} and quasars \citep{hatziminaoglou:2000} to fit the SEDs of the \otf LAE and LBG candidates. These templates correspond to different types of starburst, spirals, irregular, elliptical galaxies and quasars. %
    For fitting the SEDs, we used the available Subaru, HST and Spitzer data. We included upper limits to compute the SED. We had to provide the transmission profiles of the Spitzer/IRAC filters\footnote{Available at NASA/IPAC Infrared Science Archive: \anchor{http://irsa.ipac.caltech.edu/data/SPITZER/docs/irac/calibrationfiles/spectralresponse/}{http://irsa.ipac.caltech.edu/data/SPITZER/docs/irac/ calibrationfiles/spectralresponse/}}, which are not included in the HyperZ filter database. We did not include \otf data because the synthetic templates do not have enough resolution to fit emission lines, which may dominate over the flux in the synthetic filter. %
    With these fittings we check whether the SEDs are compatible or not with LAEs and LBGs. The results are summarized in Table~\ref{tab:interloper}. %
    There is a prudence to be taken when using HyperZ to fit the data: the number of photometric bands is rather small to obtain accurate fittings, and usually it is possible to fit the SED of various types of galaxies at different redshifts. %
%    Moreover, \citet{maraston:2006} has shown that in a sample of $z \sim 2$ galaxies, SEDs based on \citet{bruzual:2003} solutions tend to show  near-IR fluxes that are significantly lower than the Spitzer/IRAC data; this discrepancy probably gets worse at higher redshifts. %
%    And second, the synthetical spectra has a rather sparse wavelength sampling (about 25\,{\AA} in the optical and 500\,{\AA} in the infrared) that prevents to discern most spectral lines from the continuum.
    Therefore, in our case SED fitting is useful to discard candidates, but provides a moderate support for object classification, and the fits must be regarded with caution. %
    Below we present the results of fitting SEDs to the \otf LAE, LAE/LBG, and LBG candidates.
    % arranged in a similar way as in \S\ref{sec:observations}.

\subsubsection{\otf LAE candidates}

    Figure~\ref{fig:LaeSeds} shows the photometry and the SEDs of the \otf LAE candidates. 

\input{fig9}

\paragraph{\laea:}
    Candidate \laea is detected in the Subaru $V$, $i$\textquotesingle\,\,and $z$\textquotesingle\,\,filters; it lies out of the field of the HST images; it is also found in Spitzer images at 3.6--5.8\,\micron, but not in the 8\,\micron band. %
    Figure~\ref{fig:LaeSeds}a presents the \otf, Subaru and Spitzer available data for this object, and the SEDs for a starburst galaxy at $z = 3.03$ and an \oii\ interloper at $z=1.45$. The \otf photometric point shows an excess that may be ascribed to \oii\ emission. %   
% High resolution optical spectrum of starbusts galaxy 1
Figure~\ref{fig:LaeSeds}a also shows the optical high resolution SED template for a starburst with E($B-V$)$ <$ 0.1 by \citet{calzetti:1994} and \citet{kinney:1996}.\!\footnote{Templates available at \anchor{http://www.stsci.edu/hst/observatory/cdbs/cdbs_kc96.html}{http://www.stsci.edu/hst/observatory\\/cdbs/cdbs\_kc96.html}}.  This high resolution template has been moved to the starburst galaxy redshift of $z=1.45$, and it has been scaled using a \mbox{third-degree} polynomial to the low resolution starburst SED template fitted by HyperZ.  The resulting optical SED does not pretend to show the actual strengths of the emission lines, but give an idea of what should be their appearance. The $\chi^2$, in Table, \ref{tab:interloper} is not very different for the \oii\ emitter and starburst fits. 
The source is an interloper, and most probably an \oii\ emitter.

\paragraph{\laeb:}
    Photometric data and SED fitting for candidate \laeb are shown in Figure~\ref{fig:LaeSeds}b. The object is detected in Subaru filters $V$, $i$\textquotesingle\,\,and $z$\textquotesingle; it is is located out of the field of the HST images; it is found in all the Spitzer/IRAC filters. 
    
    The photometric profile using only the \otf observations (see Figure~\ref{fig:LaeCandidate}b), suggests a LAE candidate. The profile of the best starburst SED fit at $z \simeq 5.4$ supports a candidate LBG, in which case the \otf photometry point probably corresponds to a spectral feature in the absorption area around the the 9000\,\AA\ wavelength.
Therefore, object \laeb is classified as a possible LBG at $z\gtrsim 5.4$.

    % Absorption: Ca II H     3968.468
    % Emission: [Ne III]    3967.51
    %           H7          Balmer Series

\input{tableX}

\paragraph{\laec:}
    Figure~\ref{fig:LaeSeds}c shows the photometric data for candidate \laec, and the SEDs of a starburst galaxy at $z = 3.0$ and $z=1.45$. The object appears in all Subaru filters, in the field of the HST program ID5991, and is detected in the Spitzer 5.8\,\micron image. Figure~\ref{fig:LaeSeds}c is similar to Figure~\ref{fig:LaeSeds}a, with the gray line showing the optical high resolution SED template for a starburst with E($B-V$)$ <$ 0.1.%
    
    The photometric profile of the \otf observations (see Figure~\ref{fig:LaeCandidate}c) resembles the long wavelength queue of an emission line, but the HST, Subaru and Spitzer data are not consistent with a $z \simeq 6.5$ LAE candidate.
    The fit for a starburst galaxy at $z=1.45$ matches the \otf photometric point corresponding to the [\ion{O}{2}]$_{\lambda\lambda \rm{3726,3729}}$ doublet. The $\chi^2$ values for the best starburst fit and the \oii\ interloper are comparable; altogether the object is likely the later.

    % [O II]      3726.16
    % [O II]      3728.91

\input{fig10}

\subsubsection{\otf double LAE/LBG candidates}

    Figure~\ref{fig:LaeLbgSeds} shows the photometry and SED fitting for the \otf double LAE/LBG candidates. %

\paragraph{\dlla:}
    Candidate \dlla (shown in Figure \ref{fig:LaeLbgSeds}a) is detected in all of Subaru's filters; it lies out of the field of the HST images; and it is found in the Spitzer 3.6\,\micron band. Figure~\ref{fig:LaeLbgSeds}a shows the photometric data available along with the best SED fit for a starburst galaxy at $z \simeq 5.4$. The flux in the \otf synthetic filter may be dominated by the short wavelength queue of an emission line (Figure \ref{fig:DubCandidate}a), which explains that the \otf data is well above the Suprime--Cam $z$\textquotesingle\,\,band. The object is most likely an interloper.

\paragraph{\dllb:}
    Figure~\ref{fig:LaeLbgSeds}b shows the photometric data for object \dllb, and the SED of a starburst galaxy at $z \simeq 5.3$ and at $1.45$. The object is detected in the Subaru $i$\textquotesingle\,\,and $z$\textquotesingle\,\,filters; it lies outside of the HST fields; and it is also found in the Spitzer 8\,\micron band. Figure~\ref{fig:LaeLbgSeds}b is similar to Figure~\ref{fig:LaeSeds}a, with the gray line showing the optical high resolution SED template for a starburst with E($B-V$)$ <$ 0.1.%%
    
    Note that, as mentioned earlier, the centroids of the photometrical profiles of the Spitzer/IRAC images for this object are shifted 1\farcs3 with respect to their optical counterparts, and thus there is a possibility that the infrared images are misidentified. %
    Both starburst galaxy SED fits are reasonable. However, the \otf flux is well above the Suprime--Cam $z$\textquotesingle, suggesting that some of the short wavelength wing of the \oii\ line (see Figure \ref{fig:DubCandidate}b) lies in the \otf wavelength range. The object is probably an \oii\ interloper.

\subsubsection{\otf LBG candidates}
Figure~\ref{fig:LbgSeds} shows the photometry and the SEDs of the \otf LBG candidates. %
\paragraph{\lbga:}

    % Quasar SED fitting bibliography \citet{hatziminaoglou:2000}

    Images for \lbga are available through the Subaru $V$ filter, HST program ID6745 and Spitzer 3.6--5.8\,\micron bands, remains undetected in the 8\,\micron observations and lies outside the field of Subaru's $i$\textquotesingle\ and $z$\textquotesingle\ filters. %
    Figure~\ref{fig:LbgSeds}a shows the photometric data for this object (see Figure~\ref{fig:LbgCandidate}a for the set of \otf observations). The \otf photometric point could correspond with an absorption line. We also plot the best fitting of a SED, that corresponds to a young spiral galaxy at $z \simeq 2.4$. The HST optical data is incongruent with any object at $z > 4$. %
    In any case, \lbga is likely a young spiral galaxy. %

\paragraph{\lbgb:}
    Figure~\ref{fig:LbgSeds}b shows the photometric data and the SED of a starburst galaxy at $z = 5.5$. %
    The object is detected in Subaru's $i$\textquotesingle\,\,and $z$\textquotesingle\,\,filters; it lies out of the field of view of the HST images; and it is found in the Spitzer 3.6--5.8\,\micron filter but not in the $V$ and 8\,\micron bands. %
    The starburst SED fits the photometric data fairly well. The spectral profile observed with \otf (Figure \ref{fig:LbgCandidate}b) may in fact be an artefact due to the absorption region at wavelengths longer than the Lyman Break. Nonetheless, the SED fit sustains that \lbgb is a reliable LBG candidate at $z\gtrsim5.5$ rather than at 6.5 as estimated using only \otf data.

\section{Discussion} \label{sec:discuss}

    In this section we compare the results obtained with the different instrumental models applied to our simulated LAE sample, and we examine different explanations for the results of the \otf pilot observations.
\input{fig11}

\input{tab3}

\subsection{Insights from simulated data}
%% Statistics
    Table~\ref{tab:statistics} summarizes the modeled performance for detection of LAEs for the three filter models analyzed in this paper. %
    %
    % Column~1 identifies the instrument. Column~2 designates the modeled parameters studied (redshift, luminosity of the \la line, and the rest-frame EW and FWHM). Column~3 indicates the number of detections over the 5000 simulations. Columns~4, 5 and 6 correspond to the median, the first and the third quartiles of the parameter distribution of the candidates, respectively. For the present data, these statistics are adequate because most of the parameters are poorly fitted using the Gaussian distribution.
    %
    We present the statistical results using non-parametric scores (sample median and quartiles) since most of the simulated variables are poorly fitted by the Gaussian distribution.

    The main differences that can be drawn from Table~\ref{tab:statistics} correspond to the number of detections and the EW distributions of the LAE recovered in each filter. %
    % Number of candidates
    The number of LAEs recovered from our simulation using the NB921 filter in the Subaru Suprime camera and the ideal rectangular filter are similar, but they are only about 40\% of those selected using the \otf. These results are explained by the asymmetric continuum bias, the filter profiles, and by the different power of narrow-band and ultra-narrow-band survey methodologies to detect LAEs with small EW values.

    % Line EW0
    It is difficult to detect small EW LAEs. Thus, it is very interesting that the distribution of the equivalent widths at the rest-frame for \otf recovered LAEs shows a median ($\widetilde{EW}_0 = 67$\,{\AA}) that is very accurate with respect to the median of the simulations (68\,{\AA}), and that is significantly smaller than those of the ideal and NB921 filters (94 and 105\,{\AA}, respectively). This result shows that the \otf can extent the search of LAE candidates to objects with a relatively low contrast between the line and the continuum fluxes, which otherwise would be underestimated or even unnoticed in narrow-band surveys. Taking the values of the first quartile for the Subaru's EW in Table~\ref{tab:statistics}, we estimate that many LAEs with $EW(\la) < 60$\,{\AA} may remain undetected in Subaru's survey. This would explain the difference between the EW medians for LAEs at redshifts $z \simeq 5.7$ and 6.5 reported by \citet[89 and 74\,{\AA}, respectively]{kashikawa:2011}, as well as the lack of LAEs between redshifts $6.6 < z < 7.1$ with $EW(\la)$ in the range between 20 and 55\,{\AA} accounted by \citet{pentericci:2011}. In fact, these changes in the EW of LAEs are interpreted as a fast evolution of the luminosity of the \la line caused by the incomplete reionization of the Universe at redshifts $z > 6$. If LAEs at $z \simeq 6.5$ and with $EW(\la) < 60$\,{\AA} were numerous, tunable filter surveys might have a large impact on our knowledge of their LF, changing our current view.

% Lyman alpha luminosity

In addition to the EW, the mean luminosity of the \la line is also 5\% and 20\% lower for simulated LAEs detected with \otf with respect to those found using the ideal and Subaru filters, respectively.  As we mentioned above, the \otf methodology to find LAE candidates presented in this paper is able to find almost all of our simulated objects.  Now we see that \otf superior performance respect to the other instruments is not only due to an unbiased wavelength coverage, but also to differences of the line luminosity properties of the detected LAEs.

% Redshifts
    The distribution of redshifts for the \otf candidates yields the largest interquartile range of $\Delta z = 0.08$, in contrast to 0.05 for both the ideal and the NB921 filters.  This difference between the \otf and the narrow-band filters is a consequence of the respective filter transmission profiles. On one hand, narrow-band filters tend to be effective in finding LAE candidates on a rather restricted range of wavelengths of the filter bandpass, on the other \otf candidates are evenly distributed on the swept wavelength range between 9122 and 9160\,{\AA}. Moreover, the Lorentzian profile of the \otf transmission, extends some filter sensitivity to find LAEs beyond the probed wavelengths. In practice, this will translate in a excess of LAE candidates with upper redshift limits, and an excess of double LAE/LBG candidates with lower redshift limits at the blue and red borders of the set of \otf, respectively, as we have seen with the observed data presented in \S\ref{sec:observations}. %
% Luminosities and FWHMs
    Finally, the distributions of the FWHM at rest-frame do not change significantly among the different filters.

% Filter profile and detection
    The depth achieved using narrow-band filters in LAE surveys is severely limited by the filter profile. Our results using simulations agree with the analysis of Subaru's data reported by \citet{kashikawa:2011}. Thus, the output varies dramatically along the filter band, reflecting the filter response. In any case, the asymmetric continuum profile at each side of the \la line introduces a detection bias regardless the profile of the narrow-band filter. This bias is difficult to correct, as it may depend on several factors, such as the actual EW of the \la line. %
    % that make spectroscopical follow-up absolutely necessary to measure these parameters, but insufficient if there is a large fraction of undetected objects.
    Therefore, narrow-band surveys are useful to find candidates in a slim volume over a large area, but the properties that can be derived from follow-up spectroscopical observations are prone to produce biased results.

%% Cons

%    The Subaru Suprime-Cam is a mosaic of ten 2048 $\times$ 4096 CCDs that covers a 34 $\times$ 27 arcmin$^2$ field of view \citep{miyazaki:2002}. These characteristics make the Subaru Suprime-Cam a powerful instrument to perform wide field surveys. On the other hand, \otf consist of a mosaic of two detectors of 2048 $\times$ 4096 red-optimized CCDs and the instrument field of view is 8.53 $\times$ 8.67 arcmin$^2$ with small shadowed area in one side. Besides, the field is not monochromatic as the effective wavelength depends on the distance to the optical center.

    In contrast, the \otf can sample the whole range of the wavelengths of interest with a spectral resolution about 10 times larger than narrow-band filters, and thus the combined \la line and break features could be recognized, and the redshift determined to a better accuracy (Figure~\ref{fig:osiris}).  This avoids the biases introduced by the relatively large bandwidth and the extended wings of the transmission profiles of narrow-band filters. LAEs with their \la line lying between 9122 and 9260\,{\AA} are almost completely recovered (Figure~\ref{fig:histograms}).  However, \otf  also has limitations, in particular the amount of total observing time increases with the number of images, which is proportional to the desired spectral resolution.  Besides, the \otf relatively small field of view of $8.53 \times 8.67 \, \rm{arcmin}^2$ with a small shadowed area in one side, cannot compare to the wide field of Subaru Suprime-Cam ($34 \times 27 \, \rm{arcmin}^2$).  As a low resolution Fabry-Pérot spectrographa, the field of view of tunable filter instruments is limited by the dependence of the effective wavelength on the distance to the optical center (see eq.~\ref{eq:wavelengthshift}), that would spoil the desired monochromaticity in wide field images (although this effect can be compensated by wavelength scanning at the expense of telescope time).  Thus, \otf is an instrument suitable for \emph{pencil-beam} surveys spanning a relatively large volume, and for assessing the biasses produced by standard narrow-band filter surveys.

    % *******************
    % Referee 5
    % *******************
    We have looked for relationships between the simulated variables.  Aside the obvious dependences imposed by candidate selection criteria (e.g. detection and EW limits), there are no practical differences between the detected and non-detected sets of simulated data, regardless of the filter characteristics. %
    The only tiny effects that we have found involves LAES with the largest observed FWHM and located near any of the edges of the filter. Thus, objects near the blue edge, but with the \la peak off the filter range, may be still detected because part of the flux lies in the filter. On the other hand, objects near the filter red edge may be undetected because the long wavelength queue of the \la line extends beyond the transmitted wavelengths. %

    We have also investigated the effect of the number of filters used to detect LAEs.  This effect may be present in observational strategies using several narrow bands to detect LAE candidates (or in general any emission objects), such as the \otf procedure discussed in this paper.  The particular filter profile may render small changes on the results, and thus we have used ideal filter sets (rather than Gaussian or Lorentzian profiles) to characterize the several sets with different numbers of filters.  The filter passband is different for each set, thus they cover the same spectral region and effective volume.  In any case, the separation between the centers of two adjacent filter is half of the FWHM of the filter passband, as in the case of the \otf.

\input{fig13}

\input{tab5}

% Results
	% Low dependency up to 7 filters

    Table~\ref{tab:nfilt}  and Figure~\ref{fig:filtnumdep} summarize the results obtained with different sets of ideal filters.  The wavelength range ($r$) at the filters FWHM covered for all the sets is the same, and it is given by the expression:
\begin{equation}
	r = \frac{n+1}{2} \, w = 150\,\text{\AA},
\end{equation}
where $n$ is the number of filters in the set and $w$ is the filter FWHM.  We notice that the number of detections, around 2500 objects, decays slightly (3\%) from 24 to 9 filters.  However, for 5 filters there is a sharp cut-off in this number, and for the set of two filters, only 168 are detected.  For the line luminosity and redshift, the change also starts at 5 filters, with increments in their values. The rest-frame EW shows also an increment for 5 filters, but abruptly decays for 2 filter sets. The reason for this decay is that the only 168 objects detected are very luminous in both line and continuum emission. The rest-frame FWHM of the line, not shown in Figure \ref{fig:filtnumdep}, may also show an increment, but it is less significant than for the previous variables. %
	% Bimodal Distribution for 2 filters
We note that for the 2 filters set the distribution of detected objects becomes bimodal, rather than reflecting the original distribution.  This occurs because a line located at the middle of the wavelength range lies in two filters, and our detection algorithm is unable to recognize such a line when the number of filters is small. 
    
    % Best number of filters
    For \la and in general for other single and isolated emission lines, the previous analysis indicates that we might save significant telescope time, and still obtain similar results, observing through 9 filters with FWHM of 30\,\AA\ rather than 24 filters with FWHM of 12\,\AA.  In return, the lower spectral resolution may increase the number of interlopers.  However, currently the \otf resolution cannot be changed in the wavelength range of our observations.  It is worth noting that for other studies where line deblending is necessary, such as mapping H$\alpha$+[\ion{N}{2}] or the [\ion{S}{2}] doublet, it may be convenient to maintain a filter bandpass lower than 15\,\AA\ \citep[e.g.][]{lara:2010, cedres:2013}.

\input{tab4}

%% OSIRIS-TF observations
    \subsection{Excess of candidates and interlopers}

    The pilot observations obtained with the \otf embrace 5 out of 24 adjacent wavelength slices included in our complete program to find LAE candidates. Table~\ref{tab:survey} shows the values of some calculated and expected quantities. %
    %
    % Column 1 indicates the state of the observations. Column 2 indicates the number of wavelength slices planned and accomplished. Columns 3 and 4 show the irradiance and the \la luminosity lower limit. Column 5 indicates the comoving volume investigated. Finally, column 6 shows the number of expected LAEs obtained through the Schechter function [see eq.~(\ref{eq:schechter})]; because one of these numbers is less than one, these quantities have been approximated to the first decimal place to allow the discussion of the results.

% NUMBER OF LAES AND NUMBER OF CANDIDATES
    Given the LAEs and \oii\ interlopers LFs, we hardly expected to find any of these objects in the limited set of observations presented in this paper.  Regardless of this prospect, we have extracted five candidates for which \otf data are congruent with LAEs, two of them also LBG candidates (in total, there were four LBG candidates).  All of the LAE candidates were rejected after fitting broad-band SEDs, but then three of them showed as possible \oii\ interlopers, which is also a number of objects higher than expectations.  These discrepancies between the number of expected LAEs and \oii\ interlopers, and the actual number of candidates, is explained due to the low number of filters used in our current observations.  Thus, the available data expands only a very small range of wavelengths (the FWHM of the synthetic \otf is 36\,\AA\ rather than 150\,\AA\ for the complete set of 24 filters), preventing a reliable sampling of the line and the continuum to both sides of the line.  As a result, we cannot distinguish between small and large FWHM lines because only a part of the wing of a line is observed, or even discriminate between emission lines and partially observed absorption features.  These problems can be easily solved by observing through the complete set of filters.  Meanwhile, we need to rely on archive data to improve candidate selection.  Under these circumstances, even the limited number of broad-band archive data is useful to reject candidates with SED profiles not compatible with LAEs or LBGs, but insufficient to confirm the nature of the objects.
 
% Nature of the candidates
    \citet{dressler:2011} have calculated the luminosity function for \oii, [\ion{O}{3}], and H$\alpha$ interlopers for LAEs at $z \simeq 5.7$. These LFs have a sharp cut-off for luminosities $\gtrsim 3 \times 10^{41}\,\ergs$, being \oii\ emitters the most numerous of the foreground sources.  Considering a similar cut-off for $z \simeq 6.5$ interlopers, it corresponds to irradiances $> 2.2 \times 10^{-17}\,\ergscm$ for \oii at $z=1.45$, $> 9.1 \times 10^{-17}\,\ergscm$ for [\ion{O}{3}] at $z=0.82$, and $> 5.6 \times 10^{-17}\,\ergscm$ for H$\alpha$ at $z=0.39$.  Our LAE candidates listed in Table~\ref{tab:candidates} have irradiances below all these flux cut-off values, the only exception being \laea which slightly exceeds the \oii\ flux cut-off. Therefore, interlopers can enhance the number of fake candidates, in accordance with the results obtained when fitting the SEDs in \S\ref{sec:archive}.

    % *******************
    % Referee 3: interlopers
    % *******************
% FWHM of interlopers
\la observed lines at $z=6.5$ are rather wide, with FWHM around 10\,\AA. Interlopers' emission lines have observed widths that usually are well below this value. Table~\ref{tab:linefwhms} shows the observed FWHM for \la and possible interlopers. For the interlopers, a fiducial velocity for the emission lines arising from Star Formation Regions of 100\,\kms\ has been chosen. Of course, the interlopers have redshifts $z \ll 6.5$, and thus lower doppler broadening than \la. Then, all the single lines, but the unresolved \oii\ blend, have observed FWHM easily distinguishable from the \la with the OSIRIS spectral resolution. In the case of the \oii\ blend, the separation between the individual line peaks, rather than the velocities, dominates the observed FWHM.

\input{tab6}

Given the detection limit for these observations ($9 \times 10^{-18} \; \ergscm$), \oii\ interlopers at $z \simeq 1.45$ with line luminosities brighter than $L_\text{[O\,II]} > 1.166 \times 10^{41} \; \ergs$ will be detected. This yields a number of 0.2 -- 0.8 expected \oii\ interlopers in our data, depending on the luminosity function adopted \citep[respectively]{dressler:2011,takahashi:2007}. Expected numbers for the full set of planed observations are shown in Table~\ref{tab:survey}. From \citet{dressler:2011}, we expect a final efficiency of about 2/3 to find LAEs, i.e., 2 LAEs for every \oii\ interloper, when the program is fulfilled.  The OSIRIS Multi Object Spectrograph-Mode, soon available at the GTC, could be used for follow-ups if necessary.

The lines ratio \oii$\,\mathrm{\lambda 3726} /$\oii$\,\mathrm{\lambda 3729}$ between the individual lines that conform the \oii\ blend feature depends on the electronic density $N_{\rm{e}}$. Extreme cases have values $\lim_{N_{\rm{e}} \rightarrow 0} = 1.5$ and $\lim_{N_{\rm{e}} \rightarrow \infty} = 0.35$, and thus this lines ratio can be used to calculate the electronic density when it is in the range $2 < \log(N_{\rm{e}}) < 4$ \citep{pradhan:2006}. Different values of \oii\ ratios have a direct incidence on the unresolved blend FWHM measured with \otf, which makes even more difficult to distinguish between LAEs and \oii\ interlopers. There are few studies on the LF of \oii, and all of them deal with the blend as a single feature \citep{hogg:1998, gallego:2002, teplitz:2003, takahashi:2007, dressler:2011}.
    % *******************

% Reliability of the luminosity function
%    As we have shown, narrow-band surveys of LAEs may be easily biased. According to \citet{kashikawa:2006}, the resdhifts of the LAE candidates ($z \simeq 6.5$) correspond to a cosmic epoch when the reionization of the universe has not been completed. Thus it is in principle possible that there is a variation in the number density of observed LAEs in the redshift range covered by the filter for the \la line. However, the difference between expected and detected candidates cannot be explained in this way. Since our observations have been performed in a spectral region that is included inside the Subaru NB921 high response range of wavelengths, any large discrepancy between our results, and the luminosity function deduced from Subaru observations, should be attributed to the different methodologies applied. This would also account the results reported by \citet{tilvi:2010} and\citet{krug:2012}, who used ultra-narrow-band filters to search for LAEs at $z \sim 7.7$, and found about the double of LAE candidates than they expected. The possible existence of a large population of LAEs with relatively small EWs ($EW(\la) < 60$\,{\AA}), mostly ignored in narrow-band surveys, can account for such discrepancy.

% Magnification bias
    An effect to take into account is the distortion of the observed LAEs and \oii\ interlopers LF, and thus their number counts, due to the redshift dependent magnification bias \citep[e.g. ][]{bartelmann:2010, diego:2011}.  This effect increases the number of observed faint sources, but reduces their number density enlarging the angular distance between the sources.  The overall result depends on the steepness of the number-count function. For the high luminosity LAEs, such as those observed at $z \sim 6.5$, this function is steep and more sources become observable.  The situation is more complex for \oii\ interlopers, for which the high luminosity objects do not dominate the number counts.  Thus, the magnification bias may reduce the number of observed \oii\ interlopers with luminosities below $\simeq 3 \times 10^{41} \, \ergs$, and enhances the counts for more luminous sources.  In our case, the field observed with the \otf is dominated by the cluster of galaxies \cluster at redshift $z = 0.583$, that expands about $4 \times 7 \, \rm{arcmin}^{2}$. The cluster contains a gravitational lensed arc \citep{luppino:1992, tran:2005}. %
    % of a galaxy at $z = 3.146$ \citep{tran:2005}.
    The strong gravitational lens model has been discussed by \citet{verdugo:2007}, %
    % who found a bimodal mass distribution. This bimodality had been previously noticed by \citet{tran:2005} studying the galaxy population, %
    % Of the 149 spectroscopically confirmed cluster members, 113 belong to the main structure and 36 to the secondary structure; the respective velocity dispersion of the structures are 865 an 282\,$\rm{km \, s^{-1}}$. \citet{tran:2005} concluding that the cluster is in the initial stages of a merging.
    and the weak-lensing signature of the cluster was detected by \citet{hoekstra:2002}, %
    % analyzing six Hubble Space Telescope (HST) WFPC2 images. These authors
    who estimated a cluster velocity dispersion of $886 \, \rm{km \, s^{-1}}$.

    Following \citet{hildebrandt:2011}, we have used the singular isothermal sphere approximation to calculate the lens magnification as a function of the angular separation $\theta$ from the cluster center:
    \begin{equation}\label{eq:magnification}
        \mu(\theta) = \frac{\theta}{\theta - \theta_E},
    \end{equation}
    where $\theta_E$ is the Einstein radius \citep{narayan:1996} for the cluster:
    \begin{equation}\label{eq:einstein}
        \theta_E = 4 \pi \left( \frac{\sigma_v}{c} \right)^2 \frac{D_{ds}}{D_s},
    \end{equation}
    %
    % One-dimensional velocity dispersion in Narayan & Bartelmann 1996
    % Eqs. 39 and 45.
    and where $\sigma_v$ is the one-dimensional velocity dispersion \citep[we have adopted][estimate of $886 \, \rm{km \, s^{-1}}$]{hoekstra:2002}, $D_{ds}$ is the angular diameter distance from the lensing cluster to the source, and $D_s$ is the angular diameter distance from the observer to the source. We have taken a common redshift of $z=6.5$ for LAEs and $z=1.45$ for \oii\ interlopers. In our case, the Einstein radius for \cluster is $\theta_E = 17\arcsec$ and $11\arcsec$ for LAEs and \oii\ interlopers, respectively.

    The mean magnification $\avg{\mu}_{wl}$ in the weak lensing regime of the \otf field around \cluster can be calculated considering an angular separation $ 3 \theta_E \leq \theta \leq 4\arcmin $ which ranges from the weak lensing limit to the edge of the detector:
    % \[ \avg{\mu}_{wl} = 1 + \frac{\theta_E \ln (\frac{\theta - \theta_E}{2 \theta__E}}{\theta - 3 \theta_E} = 1.17 \]
    %
    \begin{equation}\label{eq:avemagn}
        \avg{\mu}_{wl} = \frac{\int^{4\arcmin}_{3 \theta_E} \mu(\theta) \; 
        \rm{d} \theta}{\int^{4\arcmin}_{3 \theta_E} \rm{d} \theta},
    \end{equation}
yielding mean magnifications of 1.17 for LAEs and 1.13 for \oii\ interlopers. Figure~\ref{fig:magnification} shows the change of the magnification with the distance to the cluster center. %
% Note that most \oii\ interloper candidates lie around 100\arcsec\ from the cluster core, where the magnification is $\mu(100\arcsec) \approx \avg{\mu}_{wl}$.  Therefore, we expect that the cluster gravitational magnification would increase the detection of LAEs.  
For the planed sample, this magnification yields an increase from 4.2 to 5.7 for expected LAE counts.  However, for the currently observed bands, the expected number of LAEs has a negligible increase of about 0.1 objects.  As discussed above, the case of \oii\ interlopers is more complicated, and depends on the actual profile of their LF.

\input{fig12}

% Cosmic variance
    A major concern with photometric searches of LAE candidates is that only a tight range of redshifts is probed, yielding small samples prone to cosmic variance due to large-scale density fluctuations. Cosmic variance accounts for deviations from the factual or expected values of the number-counts of rare phenomena, and of objects found in small volume surveys. Following \citet{trenti:2008}, we have calculated that the uncertainty on the number-counts is $\sigma_{counts} = 1$ for both LAEs and \oii\ interlopers. Moreover, \oii\ emitters are likely to show strong cross-correlation positions \citep{dressler:2011}. Therefore, we do not discard the possibility that some interloper candidates are actually \oii\ emitters.

\section{Conclusions} \label{sec:conclus}

    Narrow-band surveys allow to sample large sky areas and have been successful in finding LAE candidates. However, we have shown that the asymmetrical profile of the continuum around the \la line in LAEs yields a detection bias that affects the performance to find objects at redshifts where the \la line lies near the long wavelength edge of narrow-band photometric filters. Therefore, the Subaru survey and others that might be conducted using narrow-band filters, are prone to yield a biased LF. Besides, this methodology tends to ignore or underestimate small EW objects.

    In the case of ultra-narrow-band surveys, our simulations show that the overall performance for LAE detection using \otf is not affected by the filter transmission profile, and the LF of LAEs could be accurately calculated.  Moreover, \otf can recover simulated LAEs with \la line EWs significantly smaller than the objects recovered with narrow-band filters.  Nonetheless, tunable filters do not produce large monochromatic images, and the number of ultra-narrow-band images increases in proportion to the required spectral resolution, thus the size of the studied area is limited for practical reasons.  Therefore, both narrow-band and ultra-narrow-band surveys have different strengths and weaknesses, and thus they must be regarded as complementary strategies to study high-redshift LAEs.

    We are carrying on a program to find LAE candidates with the \otf at the GTC. Part of this program is devoted to find candidates at redshift $z \simeq 6.5$, with an strategy based on our study with Monte Carlo data.  We have already performed pilot observations of five sets of images with the GTC and the \otf instrument at adjacent wavelengths, which is a fraction of the 24 wavelength slices that we plan to observe. The \otf images are separated by wavelength steps of 6\,{\AA}, and have a bandpass of 12\,{\AA}, covering a wavelength range of about 36\,{\AA}. The total exposure time in each wavelength is 630\,s, rendering a detection level of $9 \times 10^{-18} \, \ergscm$, which is about two times less sensitive than that reported for the Subaru LAE surveys. Available Subaru, HST/WFPC2 and Spitzer/IRAC archive data have been employed to build SED models. Because of the limited wavelength range of \otf observations, these models have been very helpful to reject LAE and LBG candidates and to identify interlopers, although the low number of bands used in the fits prevents accurate object classification and redshift estimate.

    We have calculated the number of expected LAEs in the \otf field with our observational conditions.  For this purpose, we have taken into account the weak lensing effect introduced by a nearby cluster of galaxies, and the effect of the cosmic variance.  Thus, we expected no more than one possible LAE and \oii\ interloper showing up in our data.  Actually, we have identified three possibly \oii\ interlopers, and one LBG candidate.  The possible overabundance of \oii\ interlopers in the field might be a result of cross-correlation positions \citep{dressler:2011}.  In any case, these results support the capabilities of \otf to perform an accurate survey of emission-line objects.
    
%    If the number of \otf confirmed LAEs results significantly larger than those expected from their currently accepted LF, it would be an evidence of the presence of an underestimated population of LAEs at redshift $z \simeq 6.5$ with $EW(\la) < 60$\,{\AA}. These results support the capabilities of \otf to perform an accurate LAE survey.

    \paragraph{Future work}
    We plan to complete the set of \otf 24 wavelength slices to extract a sample of high-redshift candidates, namely LAEs, LBGs and high-redshift interlopers. The sample will be studied to confirm the nature of the objects using OSIRIS Multi-Object Spectrograph when available.

\acknowledgments

    This research has been partially funded by the UNAM-DGAPA-PAPIIT IN110013 Program. %
	J.A.D. and M.A.D. are grateful for support from CONACyT grant CB-128556. %
    J.A.D. is grateful for support from grant SAB2010-0011 awarded by the Spanish MIED through the ``Programa Nacional de Movilidad de Recursos Humanos'' included in the Plan Nacional de I-D+i 2008-2011. %
    T.V. acknowledges support from CONACYT grant 165365 through the program ``Estancias posdoctorales y sab\'{a}ticas al extranjero para la consolidaci\'{o}n de grupos de investigaci\'{o}n.'' %
    This work has been partially funded by the Spanish Ministry of Science and Innovation (MICINN) under the Consolider-Ingenio 2010 Program grant CSD2006-00070: First Science with the GTC (\url{http://www.iac.es/consolider-ingenio-gtc}),   AYA2011-29517-C03-01, and AYA\-2011-29517-C03-02. %
    Observations presented in this paper were made with the Gran Telescopio Canarias (GTC), instaled in the Spanish Observatorio del Roque de los Muchachos of the Instituto de Astrof\'{\i}sica de Canarias, in the island of La Palma.
    The authors are thankful to the anonymous referee for the critical and constructive suggestions.

{\it Facilities:} \facility{GTC (OSIRIS)}.

%\appendix

%\bibliographystyle{apj}	 % (uses file "plain.bst")
%\bibliography{jdo}		 % expects file "jdo.bib"

\input{ref.tex}
\end{document}

%% file: fig2.tex
    \begin{figure}[t]
      % LAE 1 at z = 6.51361
      % LAE 2 at z = 6.60570
      \includegraphics[width=\linewidth]{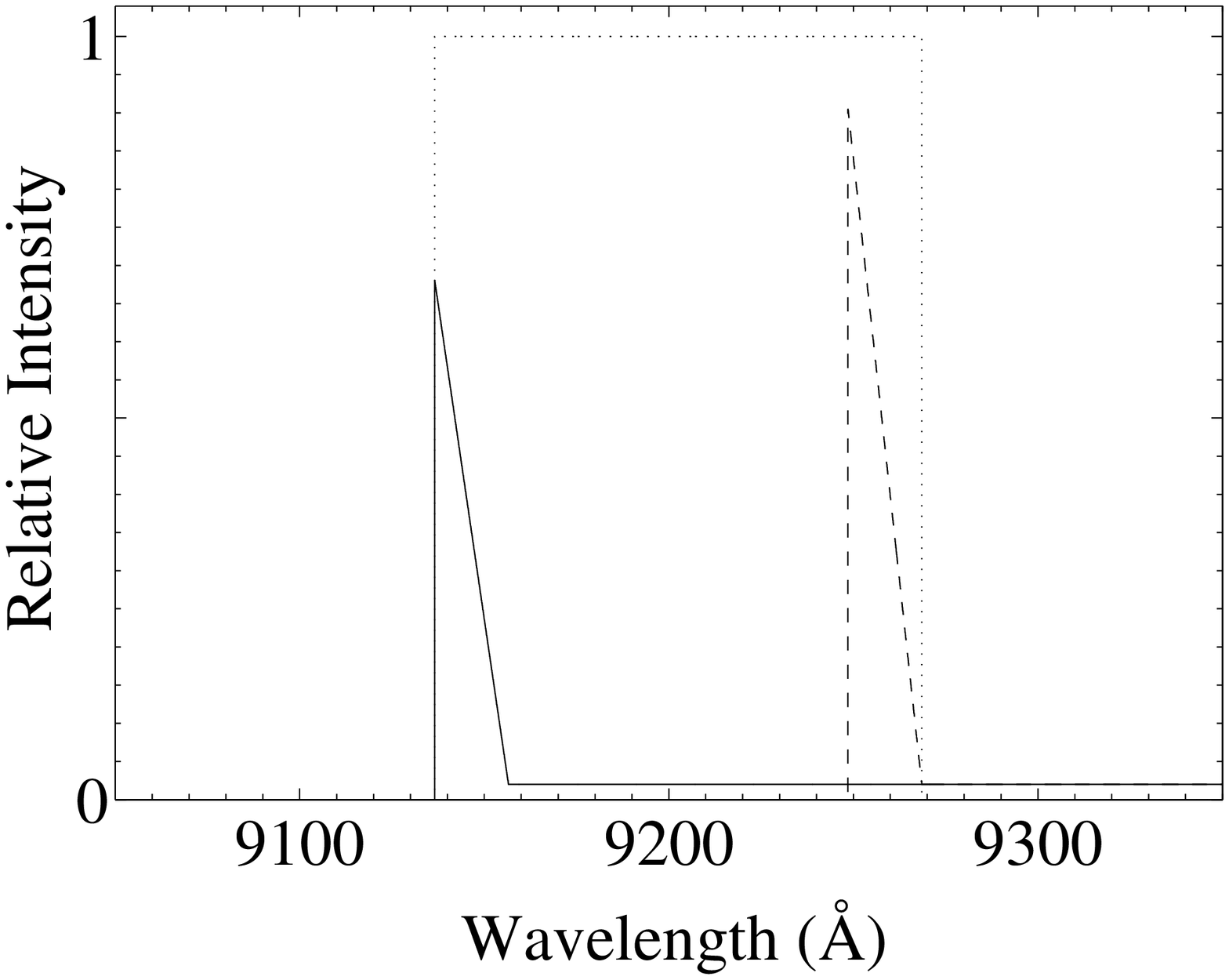}\\
      \caption{An example of the effect of the asymmetric continuum in the detection of LAEs. The amount of the continuum inside the ideal rectangular filter (dotted line) depends on the position of the \la break. In the extreme cases shown in this diagram, the continuum expands on the whole wavelength range of the filter for objects at the lowest redshift (solid line) or barely in the region covered by the \la line for objects at the highest redshift (dashed line). Both objects share the same continuum level and the observed FWHM of the \la lines is 10\,{\AA}. For the object at the lower redshift, the contribution of the \la line is 2.5 that of the continuum in the range of the filter. For detecting the highest redshift object with the same total signal, the \la line must be approximately 34\% more intense than the line of the lower redshift object.}\label{fig:bias}
    \end{figure}

%% file: fig3.tex
    \begin{figure}[th!]
      \includegraphics[width=\linewidth]{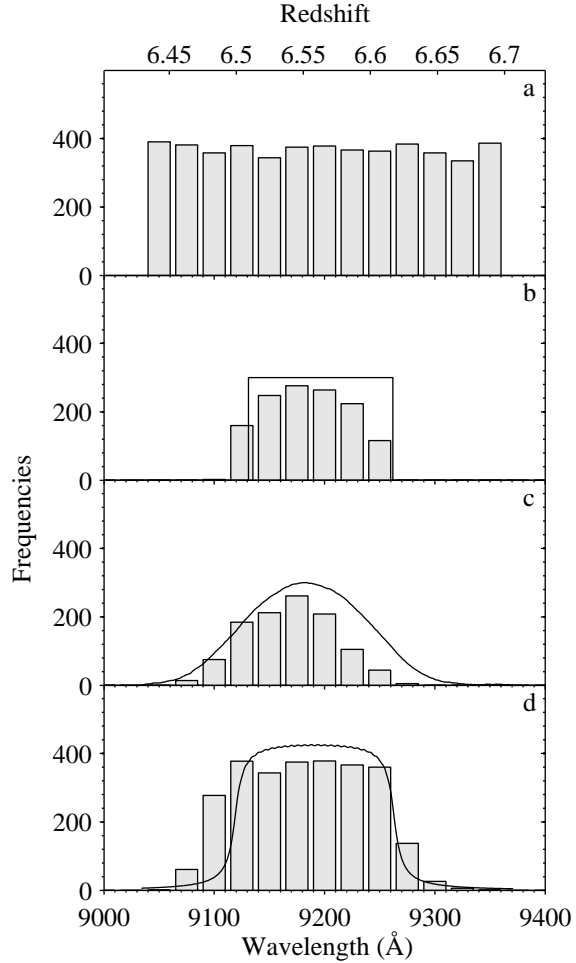}\\
      \caption{Number of simulations and detections. Panel (a) shows the absolute frequency per wavelength bin of the simulated population of LAEs. Panels (b) and (c) depict the number of detections computed for an ideal rectangular filter and for the Subaru NB921 filter; both panels show the transmission profile for the respective filter superposed. Panel (d) renders the number of detections computed for the \otf, along with the transmission profile of the wide filter (solid line) built using the band synthesis technique (see text).}\label{fig:histograms}
    \end{figure}

%% file: fig4.tex
    \begin{figure}[t!]
      \includegraphics[width=\linewidth]{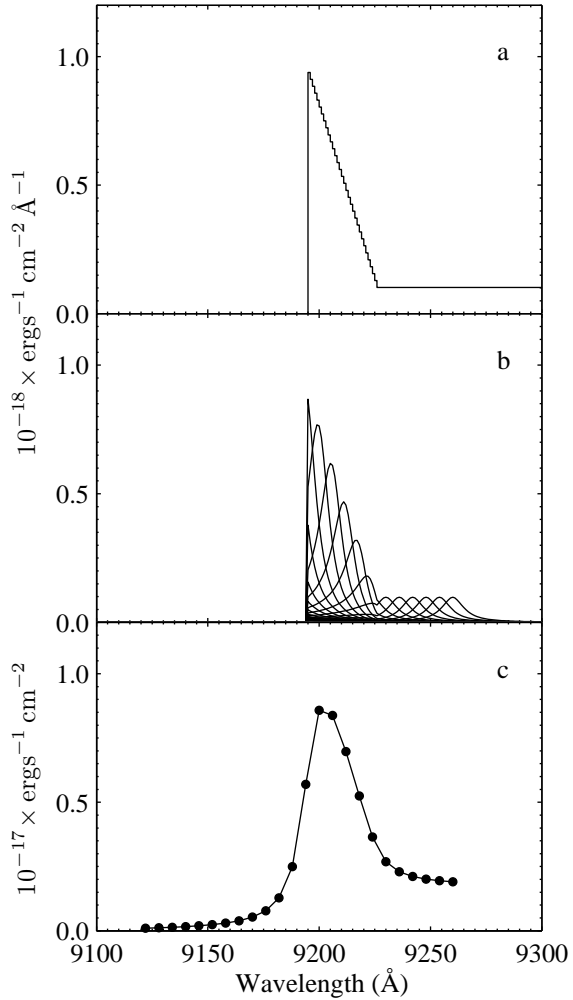}\\
      \caption{\otf output. Panel (a) shows the spectral flux density of a LAE at $z=6.6$ simulated spectrum. Panel (b) shows the set of \otf transmitted spectra for this simulation.
      % shows the LAE spectrum transmitted through each \otf considered in our simulations.
      Panel (c) renders the photometric output for each \otf, which is the signal integrated over each filter. Note that the units and the scale of the last plot are different from the two previous diagrams.}\label{fig:osiris}
    \end{figure} 

%% file: tab1.tex
    \begin{table*}
    \centering
    \caption{High redshift \otf candidates}\label{tab:candidates}
    %\small
    \begin{tabular}{cr@{:}r@{:}lr@{:}r@{:}lrr@{ $\pm$ }lcc}
    % HST-WFPC2 filter infomation at:
    %   http://www.stsci.edu/hst/wfpc2/documents/wfpc2_filters_archive.html
      \tableline \tableline \noalign{\smallskip} %
      \mr{ID\tn{a}} & \multicolumn{3}{c}{RA\tn{b}} & \multicolumn{3}{c}{DEC} & \mr{Redshift}   & \mc{\mr{Irradiance\tn{c}}} &     \mc{Candidates}        \\
              & \multicolumn{3}{c}{\small hh:mm:ss.s} & \multicolumn{3}{c}{\small dd:mm:ss} &                 & \mc{ }                     & \otf only  & \otf \& SED   \\
      \tableline \noalign{\smallskip} %
     \laea    & 20 & 56 & 23.8 & -4 & 40 & 07 & 6.494           & 4.01   & 0.09              & LAE        & Interloper / \oii\,\,emitter          \\
     \laeb    & 20 & 56 & 25.0 & -4 & 37 & 07 & 6.498           & 1.05   & 0.09              & LAE        & LBG    \\
     \laec    & 20 & 56 & 23.8 & -4 & 37 & 02 & $\lesssim$6.494 & 1.12   & 0.09              & LAE        & Interloper / \oii\,\,emitter   \\
     \dlla    & 20 & 56 & 38.1 & -4 & 40 & 04 & $\gtrsim$6.448  & 1.49   & 0.09              & LAE/LBG    & Interloper           \\
     \dllb    & 20 & 56 & 26.4 & -4 & 37 & 14 & $\gtrsim$6.512  & 1.67   & 0.09              & LAE/LBG    & Interloper / \oii\,\,emitter       \\
     \lbga    & 20 & 56 & 16.7 & -4 & 37 & 53 & 6.513           & 1.00   & 0.09              & LBG        & Young spiral galaxy    \\
     \lbgb    & 20 & 56 & 30.6 & -4 & 37 & 37 & 6.502           & 1.82   & 0.09              & LBG        & LBG          \\
%*=strong	case
%IRAC AORKEY=18626048 ;	P.I.	TRAN;	YEAR=2007-11-13;	R.A.	 314.095833 DEC.	-4.627778
      \tableline
    \end{tabular}
      \tablenotetext{a}{Candidate identifier.}
      \tablenotetext{b}{Epoch J2000.}
      \tablenotetext{c}{In units of $10^{-17}\ergscm$.}
    \end{table*}

%% file: fig5.tex
\begin{figure*}[p]
  \thisfloatpagestyle{empty}
  \def\mywidth{0.95\linewidth}
  \def\myrise{1.2cm} % No rotation
  \def\myhs{\hspace{0.57in}}
  \newcommand{\myrot}[1]{#1}
 % \def\myrise{0.8cm} % Rotation
 % \newcommand{\myrot}[1]{\begin{sideways}#1\end{sideways}}
 % \begin{tabular*}{0.97\textwidth}{@{\extracolsep{\fill}}l*{6}{c}}
 %     & Deep & 9122\,{\AA} & 9128\,{\AA} & 9134\,{\AA} & 9140\,{\AA} & 9146\,{\AA} \\
  \begin{tabular*}{\textwidth}{ll}
     & \hspace{0.30in} Field \pa \myhs 9122\,{\AA} \myhs 9128\,{\AA} \myhs 9134\,{\AA} \myhs 9140\,{\AA} \myhs 9146\,{\AA} \\
  \end{tabular*}
  \begin{tabular*}{\textwidth}{@{}l@{}l@{}}
    \raisebox{\myrise}{\myrot{\laea}} & \includegraphics[width=\mywidth]{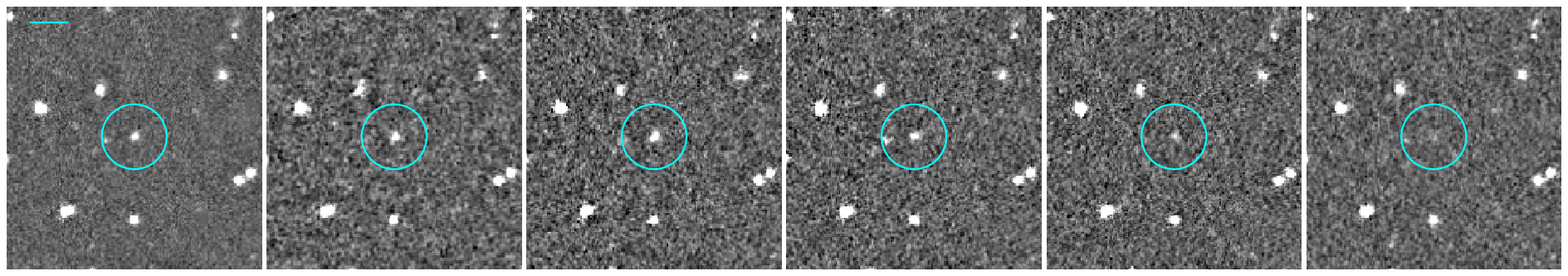} \\
    \raisebox{\myrise}{\myrot{\laeb}} & \includegraphics[width=\mywidth]{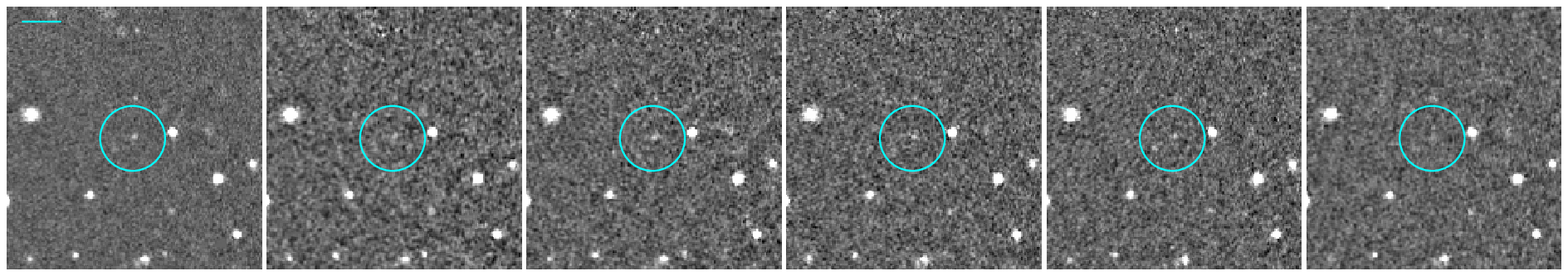} \\
    \raisebox{\myrise}{\myrot{\laec}} & \includegraphics[width=\mywidth]{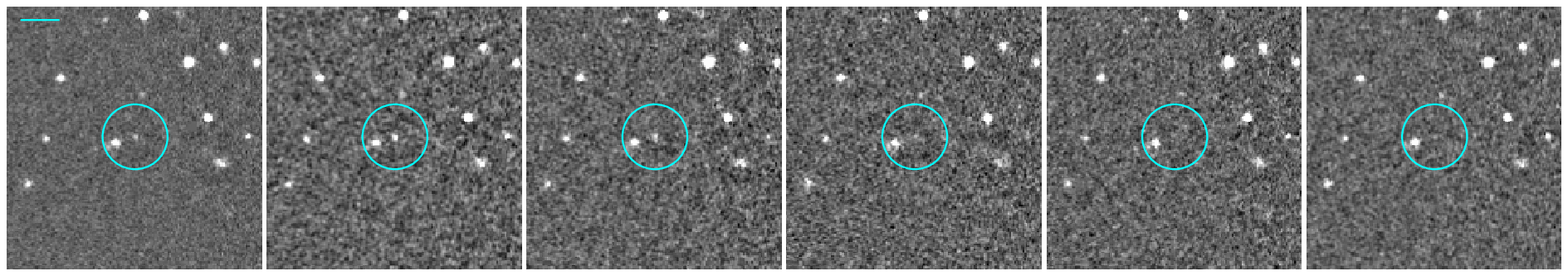} \\
    \raisebox{\myrise}{\myrot{\dlla}} & \includegraphics[width=\mywidth]{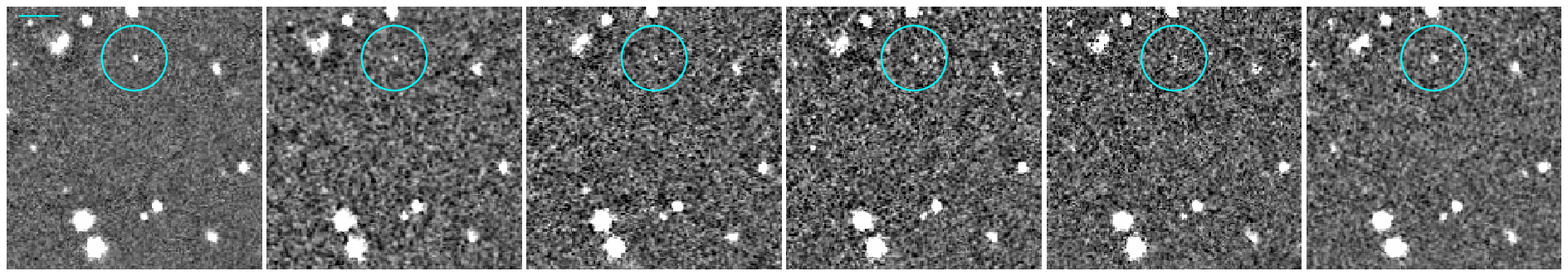} \\
    \raisebox{\myrise}{\myrot{\dllb}} & \includegraphics[width=\mywidth]{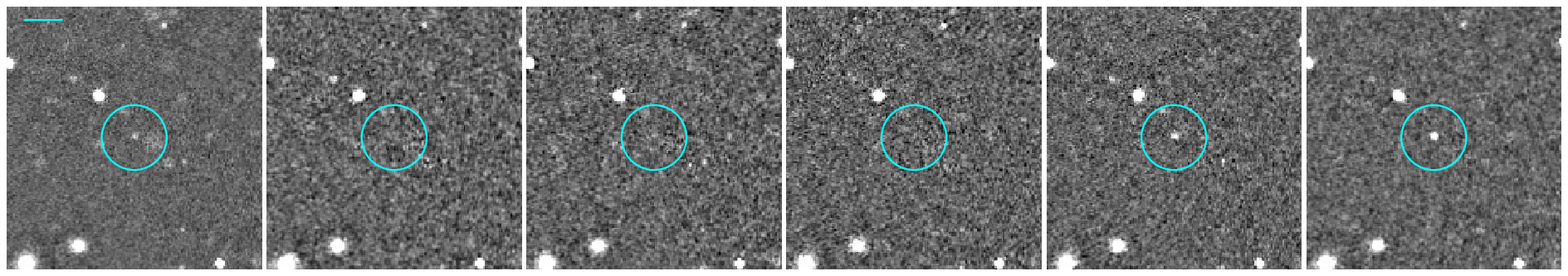} \\
    \raisebox{\myrise}{\myrot{\lbga}} & \includegraphics[width=\mywidth]{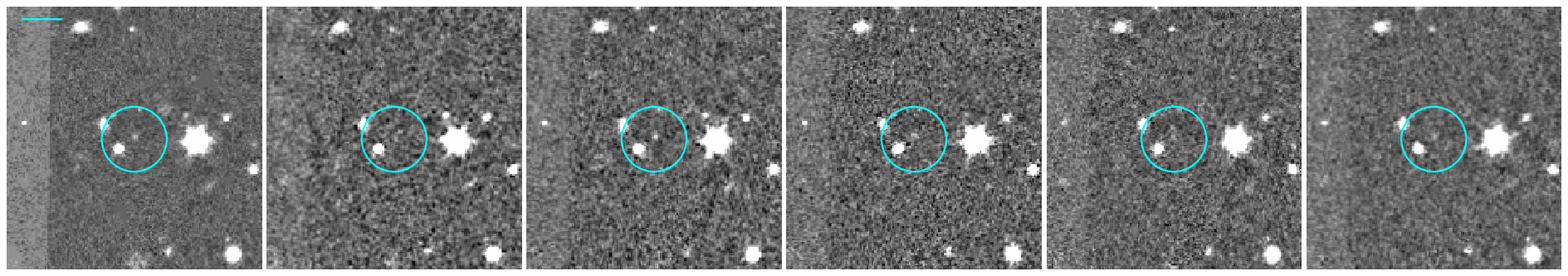} \\
    \raisebox{\myrise}{\myrot{\lbgb}} & \includegraphics[width=\mywidth]{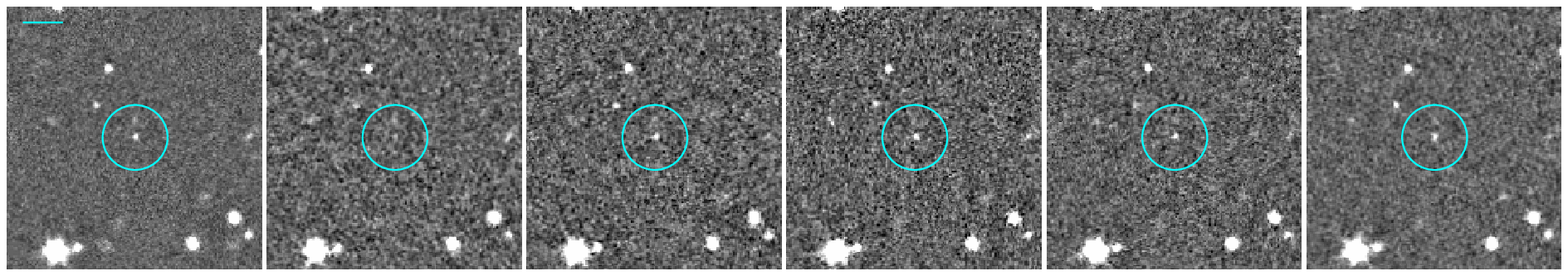} \\
  \end{tabular*}
  \caption{\otf observations. Column 1 shows the field obtained adding five individual images shown in the columns 2--6 labeled by the central wavelength of the tuned filter. The horizontal line at the top left of the fields marks a 6\arcsec\ scale. Each row corresponds to a different LAE or LBG candidate, identified by the labels at left. The objects lie in the center of the identifying circles.}\label{fig:observations}
\end{figure*}

\floatpagestyle{plain} 

%% file: fig6.tex
%    Redshift of candidate c292 is z =  6.494
%    	 Signal is 4e-017 x erg s^{-1} cm^{-2}
%    	 Simulation 322 	 z(322) = 6.5011
%        File paper/jdo3.eps has been saved
%
%    Redshift of candidate c303 is z =  6.448
%    	 Signal is 1.5e-017 x erg s^{-1} cm^{-2}
%    	 Simulation 3347 	 z(3347) = 6.5253
%        File paper/jdo4.eps has been saved
%
%    Redshift of candidate c745 is z =  6.498
%    	 Signal is 1e-017 x erg s^{-1} cm^{-2}
%    	 Simulation 4918 	 z(4918) = 6.5018
%        File paper/jdo5.eps has been saved
%
%    Redshift of candidate c801 is z =  6.492
%    	 Signal is 1.8e-017 x erg s^{-1} cm^{-2}
%    	 Simulation 4918 	 z(4918) = 6.5018
%        File paper/jdo6.eps has been saved
%
%    Redshift of candidate c889 is z =  6.512
%    	 Signal is 1.7e-017 x erg s^{-1} cm^{-2}
%    	 Simulation 2064 	 z(2064) = 6.5178
%        File paper/jdo7.eps has been saved
%
%    Redshift of candidate c913 is z =  6.498
%    	 Signal is 1.1e-017 x erg s^{-1} cm^{-2}
%    	 Simulation 959 	 z(959) = 6.5016
%        File paper/jdo8.eps has been saved
%
%    Redshift of candidate c928 is z =  6.494
%    	 Signal is 1.1e-017 x erg s^{-1} cm^{-2}
%    	 Simulation 2478 	 z(2478) = 6.5
%        File paper/jdo9.eps has been saved

    \begin{figure}
      \centering
      \begin{tabular}{c}
%  Redshift of candidate c292 is z =  6.494
%	 Simulation 322 	 z(322) = 6.5011
      \includegraphics[width=0.45\textwidth]{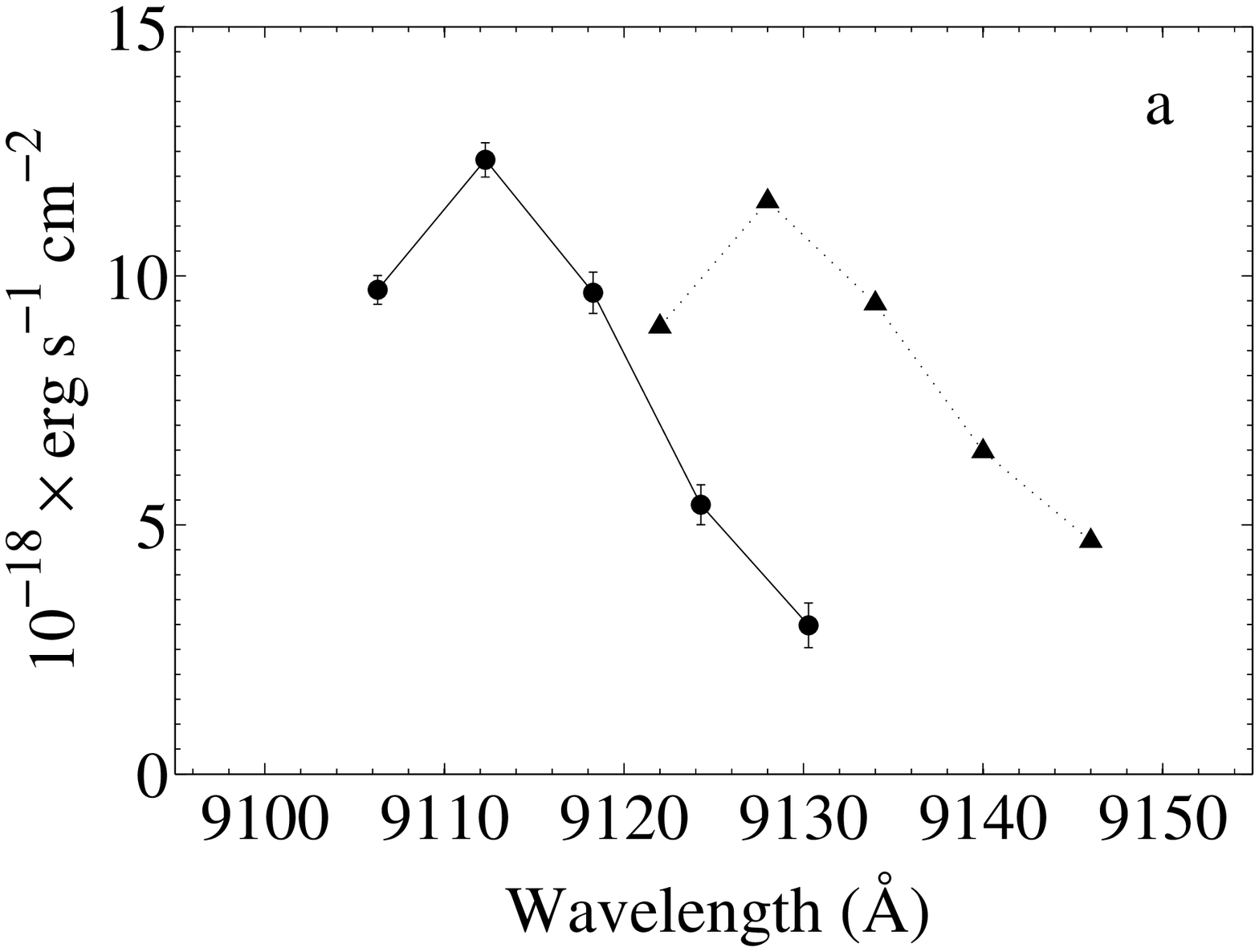} \\
%  Redshift of candidate c913 is z =  6.498
%  	 Simulation 959 	 z(959) = 6.5016
      \includegraphics[width=0.45\textwidth]{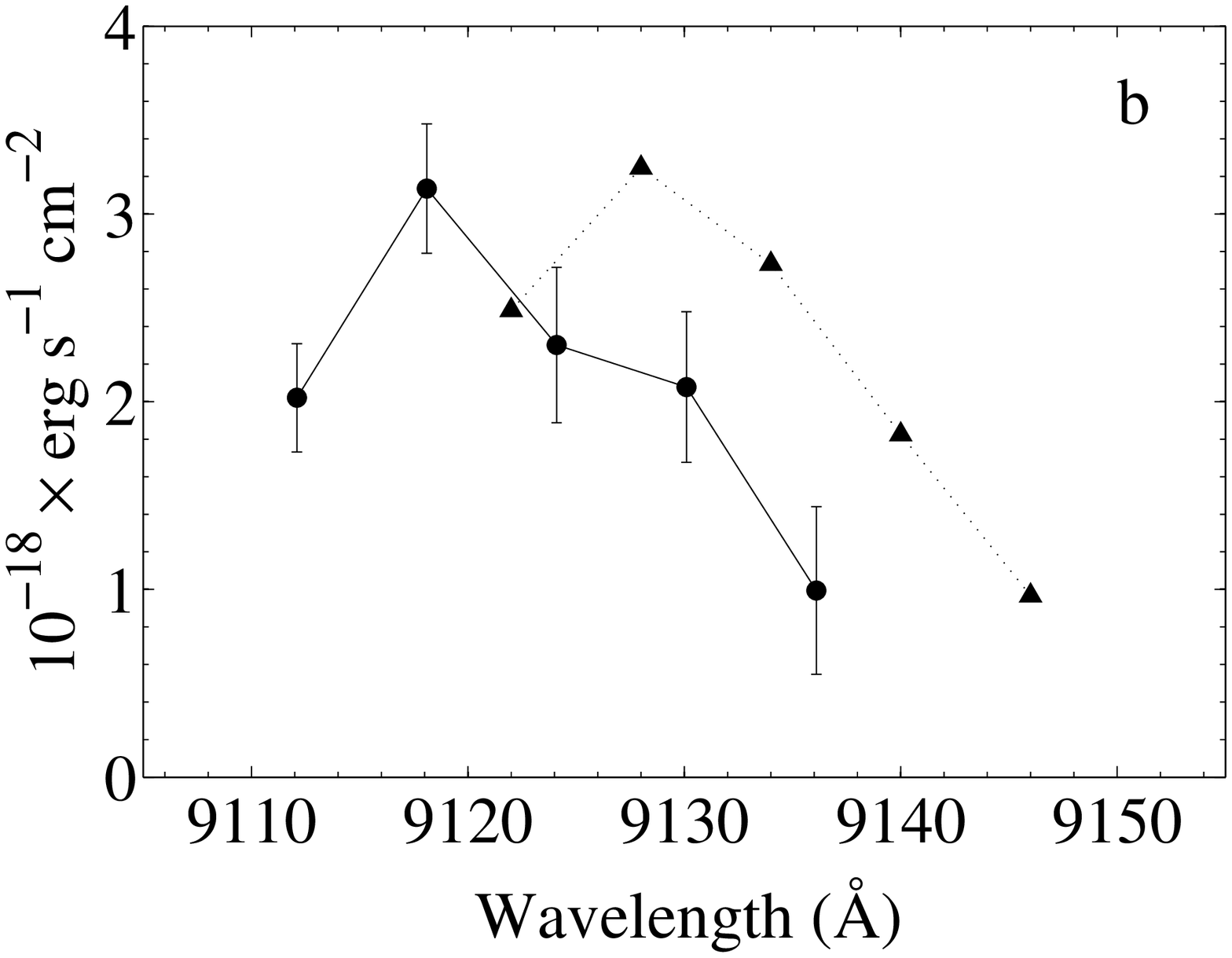} \\
%  Redshift of candidate c928 is z =  6.494
%  	 Simulation 2478 	 z(2478) = 6.501
      \includegraphics[width=0.45\textwidth]{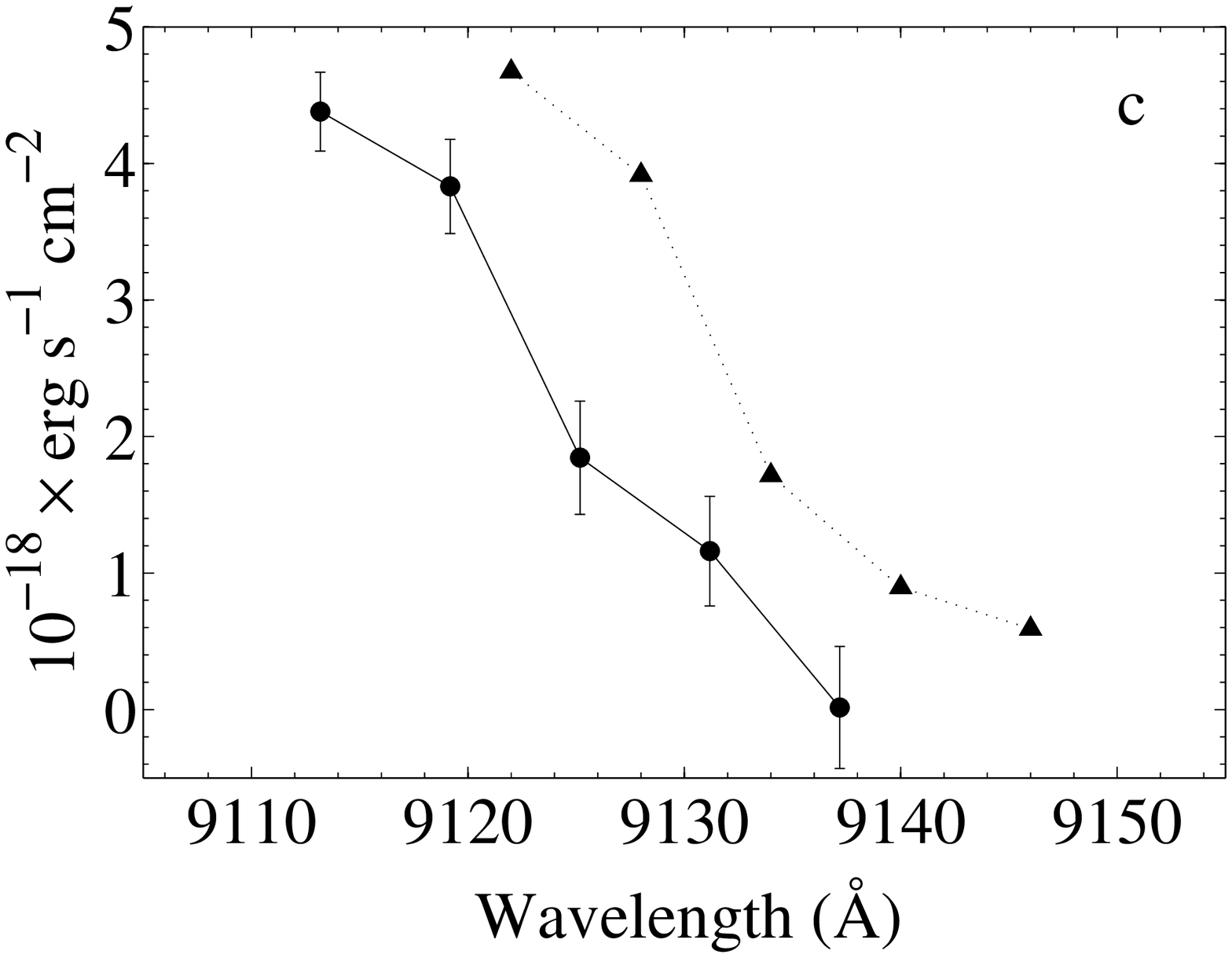}\\
      \end{tabular}
      \caption{\otf LAE candidates. Panels (a), (b) and (c) correspond to candidates \laea, \laeb, and \laec, respectively. Circles connected by solid lines depict \otf photometrical data obtained at the GTC. Triangles linked by dotted lines show examples extracted from our simulations that resemble the real data. 
Photometrical data are shifted in wavelength with respect to the simulations due to the wavelength dependency of the \otf with the distance to the optical center.
}\label{fig:LaeCandidate}
    \end{figure} 

%% file: fig7.tex
    \begin{figure*}
      \centering
      \begin{tabular}{lr}
%  Redshift of candidate c303 is z =  6.415
%    Simulation 3347 	 z(3347) = 6.5253
      \includegraphics[width=0.45\textwidth]{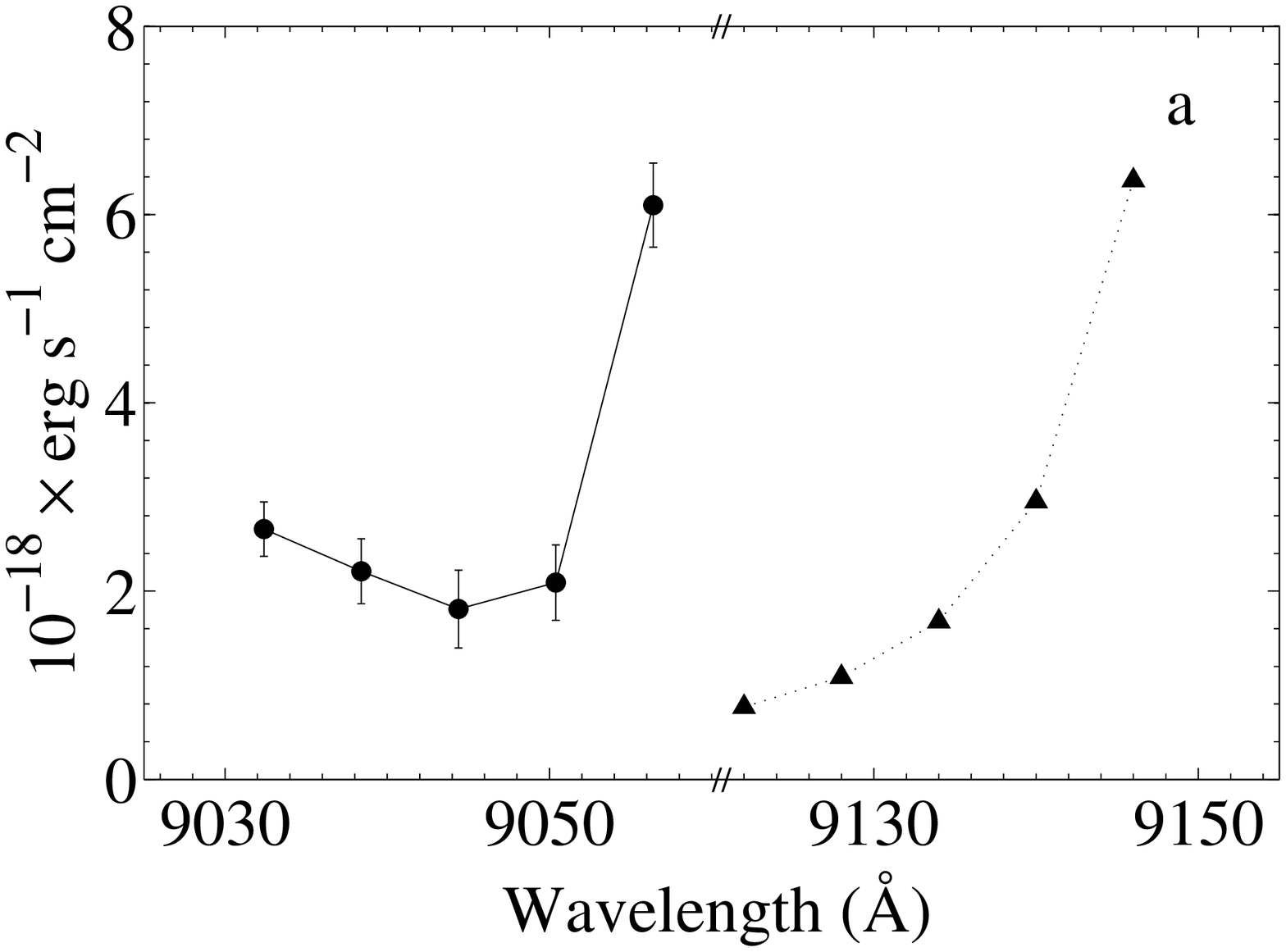} &
%  Redshift of candidate c889 is z =  6.508
%  	 Simulation 2064 	 z(2064) = 6.5178
      \includegraphics[width=0.45\textwidth]{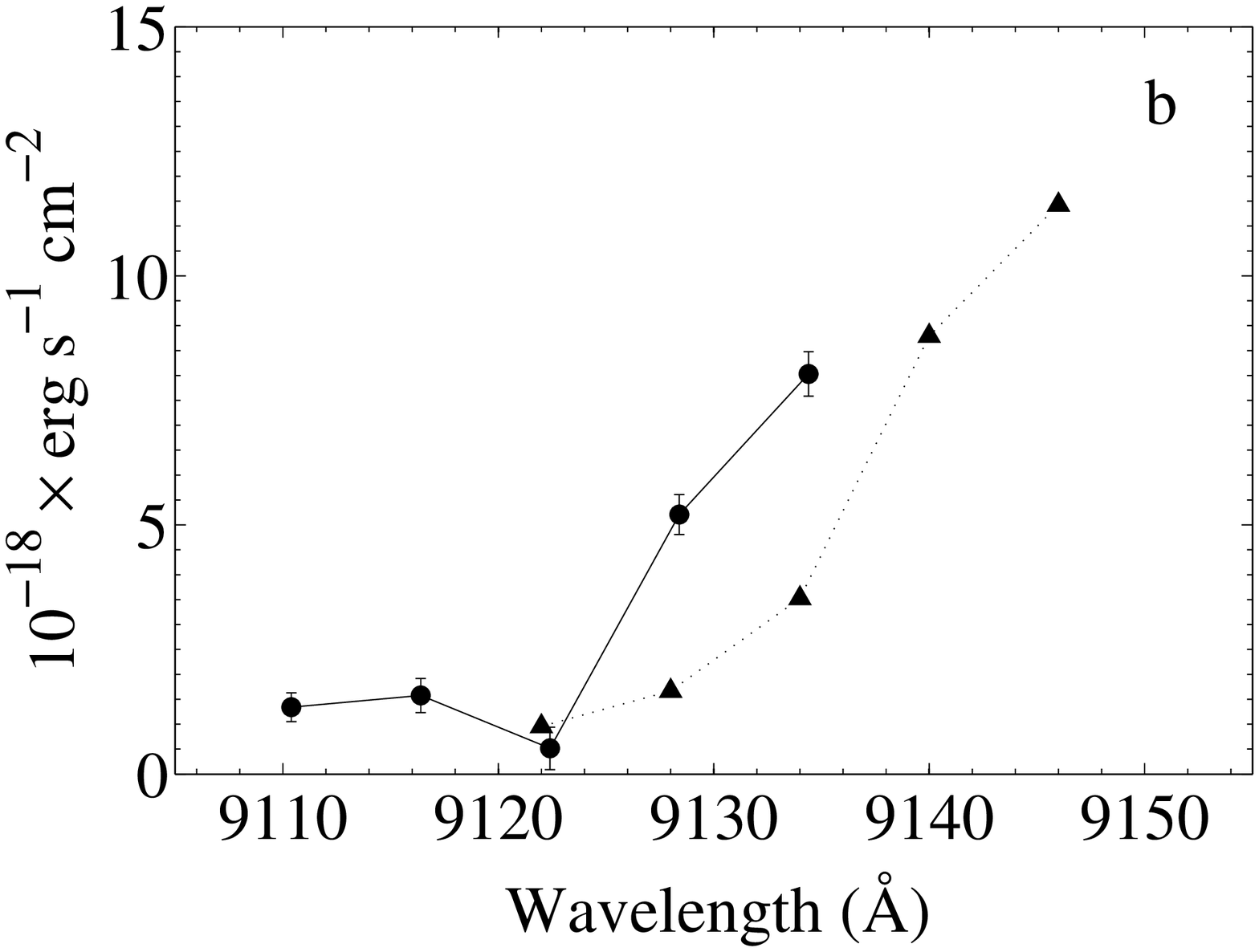}\\
      \end{tabular}
      \caption{\otf LAE or LBG candidates. The same as in  Figure~\ref{fig:LaeCandidate}, but the spectral profiles are  compatible to both LAEs and LBGs. Panels (a), and (b) correspond to candidates \dlla, and \dllb, respectively.}\label{fig:DubCandidate}
    \end{figure*} 

%% file: fig8.tex
    \begin{figure*}
      \centering
      \begin{tabular}{lr}
%  Redshift of candidate c745 is z =  6.494
%    Simulation 4918 	 z(4918) = 6.5018
      \includegraphics[width=0.45\textwidth]{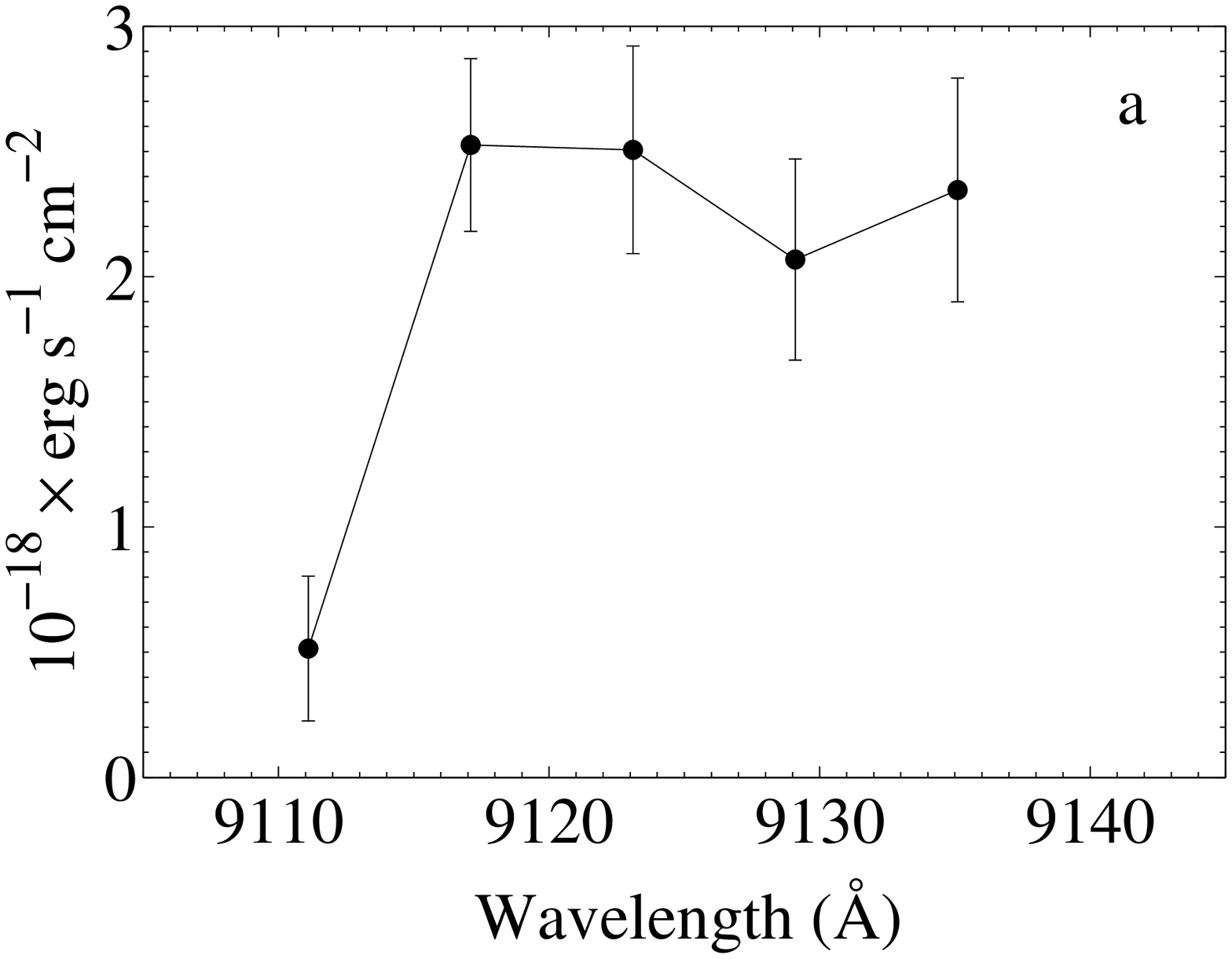} &
%  Redshift of candidate c801 is z =  6.484
%    Simulation 4918 	 z(4918) = 6.5018
      \includegraphics[width=0.45\textwidth]{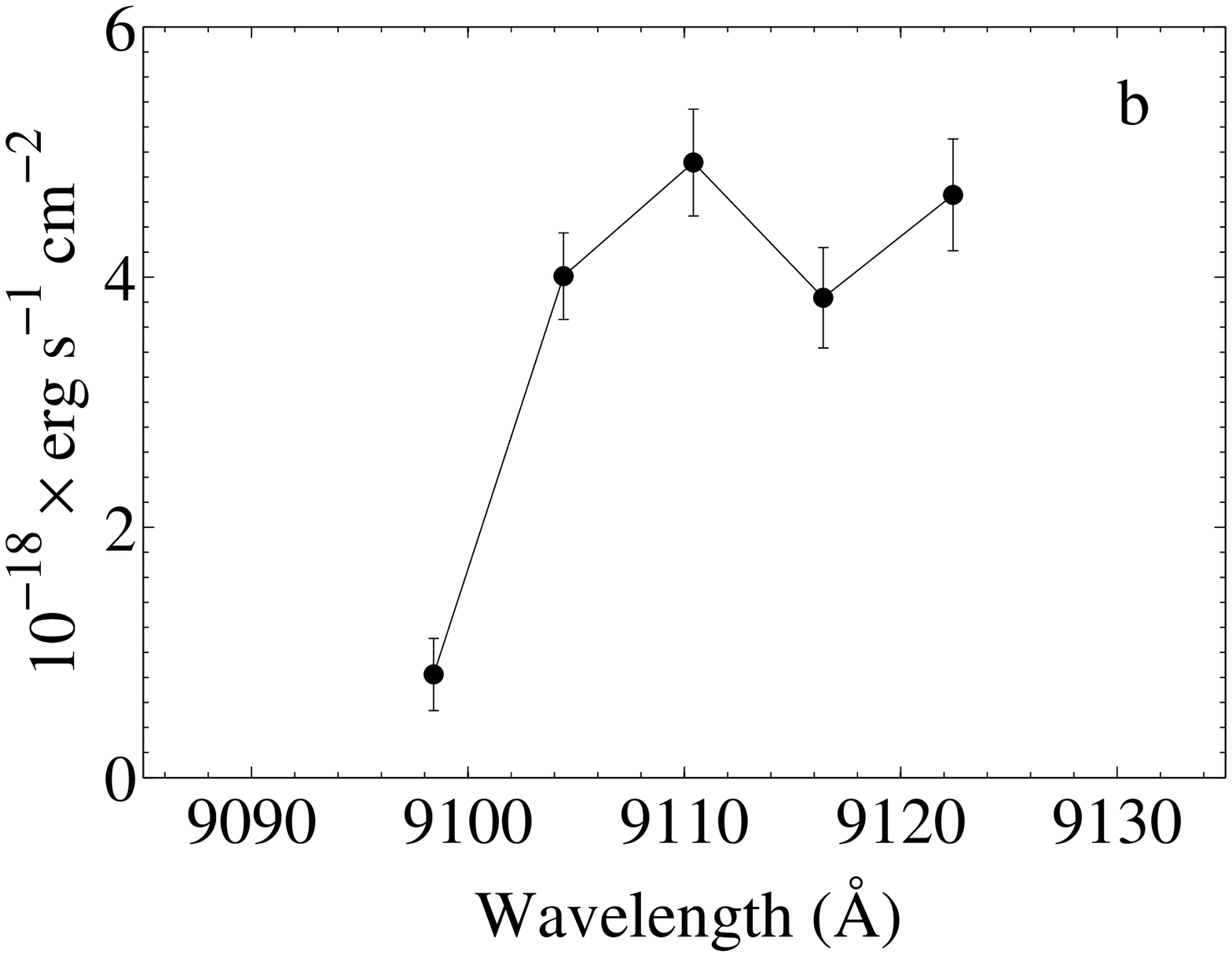}\\
      \end{tabular}
      \caption{\otf LBG candidates. Panels (a), and (b) correspond to candidates \lbga, and \lbgb, respectively. Circles with error bars linked by solid lines depict photometrical data obtained with the \otf at the GTC. The spectral profiles are compatible with LBGs.}\label{fig:LbgCandidate}
    \end{figure*} 

%% file: tab2.tex
\begin{table*}[t]
\centering
\caption{Candidate Fluxes}\label{tab:fluxes}
\begin{tabular}{@{}lr@{$\,\pm\,$}lr@{$\,\pm\,$}lr@{$\,\pm\,$}lr@{$\,\pm\,$}lr@{$\,\pm\,$}lr@{$\,\pm\,$}lr@{$\,\pm\,$}l@{}}
 \tableline \tableline \noalign{\smallskip} %
 FILTER  & \mc{\laea} &  \mc{\laeb} &  \mc{\laec} &  \mc{\dlla} &  \mc{\dllb} &  \mc{\lbga} & \mc{\lbgb} \\
 \tableline \noalign{\smallskip} %
 Suprime $V$ 	& 26		&  2		& \mc{$<0.70$}	& 4		& 1		& 8.3	& 0.7	& \mc{$\cdots$}	& 36		& 1		& \mc{$\cdots$}	\\
 F606W      	& \mc{$\cdots$}	& \mc{$\cdots$}	& \mc{$\cdots$} 	& \mc{$\cdots$}	& \mc{$\cdots$}	& 41		& 7		& \mc{$\cdots$}	\\
 F702W      	& \mc{$\cdots$}	& \mc{$\cdots$}	& 1.5	& 0.2 	& \mc{$\cdots$}	& \mc{$\cdots$}	& \mc{$\cdots$}	& \mc{$\cdots$}	\\
 Suprime $i'$ 	& 7.3	& 0.3	& 7.5	& 0.2	& 1.3	& 0.1	&11.0	& 0.3	& 4.0	&0.2		& \mc{$\cdots$}	& 9.7	& 0.2	\\
 F814W      	& \mc{$\cdots$}	& \mc{$\cdots$}	& 0.5	& 0.2	& \mc{$\cdots$}	& \mc{$\cdots$}	& 9.0	& 0.6	& \mc{$\cdots$}	\\
 \otf          	& 38		& 1		& 18 	& 2		& 10		& 1	 	& 22 	& 2		& 12.0	& 0.8	& 15		& 1		& 20.9	& 0-3	\\
 Suprime $z'$	& 7.3	& 0.4	& 21.0	& 0.6	& 0.7	& 0.4	& 16		& 2		& 4.0	& 0.4	& \mc{$\cdots$}	& 17.0	& 0.3	\\
 3.6 $\mu$m 	& 0.55	& 0.03	& 10.34	& 0.04	& \mc{$<0.15$}	& 0.41 & 0.06		& \mc{$<0.15$}	& 1.86	& 0.03	& 0.44	& 0.02	\\
 4.5 $\mu$m 	& 0.47	& 0.03 	& 6.37	& 0.03	& \mc{$<0.08$}	& \mc{$<0.08$}	& \mc{$<0.08$}	& 1.15	& 0.03	& 0.24 	& 0.03	\\
 5.8 $\mu$m 	& 0.80	& 0.07 	& 3.47	& 0.07	& 0.69 & 0.07		& \mc{$<0.14$}	& \mc{$<0.14$}	& 0.77	& 0.07	& 0.36	& 0.07	\\
 8.0 $\mu$m 	& \mc{$<0.17$}	& 1.73	& 0.08	& \mc{$<0.17$}	& \mc{$<0.17$}	& 0.44 & 0.09		& \mc{$<0.17$}	& \mc{$<0.17$}	\\
 \tableline
\end{tabular}
\tablecomments{All the fluxes in units of 10$^{-19}$\,\ergscma.}
\end{table*}

%% file: fig9.tex
    \begin{figure}
      \centering
      \includegraphics[width=\linewidth]{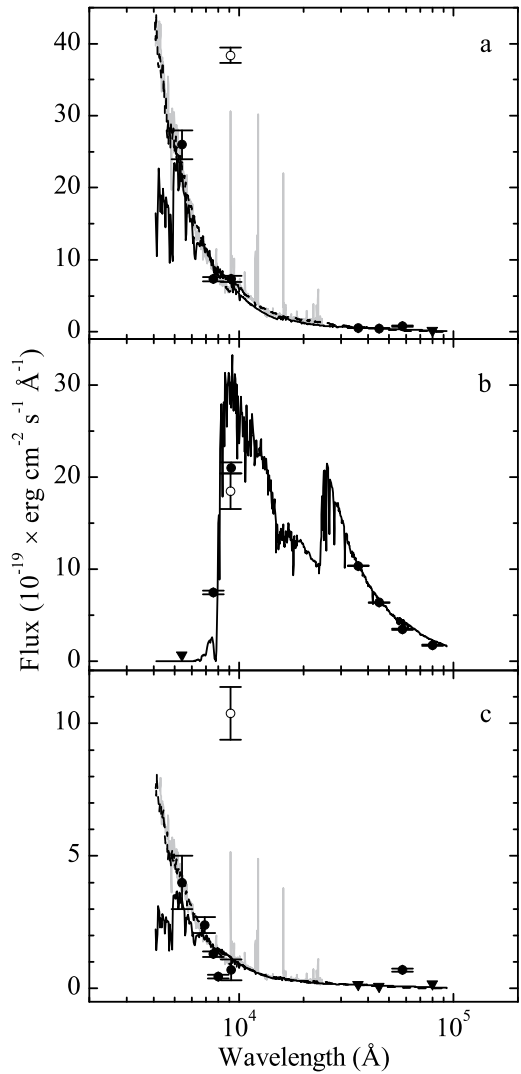}\\
      \caption{
%SED fittings for \otf LAE candidates. \otf (hallow circle), Subaru, HST, and Spitzer (filled circles for all three) fluxes are plotted along with starburst galaxies SEDs calculated with HyperZ. (a) Candidate \laea: the SED was calculated at $z=3.03$ (solid line); the dashed line is a fit at $z=1.45$. The possible excess flux excess around 9000\,{\AA} suggests the presence of \oii\ emission. The gray line shows a scaled high--resolution SED template for a starburst galaxy by \citet{kinney:1996}. The best fit suggests an \oii\ interloper. (b) Candidate \laeb was fitted at $z\simeq5.4$ with a profile similar to an LBG, suggesting that the object may be an LBG at $z\gtrsim5.4$ or an interloper. (c) Candidate \laec: the photometric data was fitted with at $z=3.0$ (solid line) and at $z=1.45$ (dashed line).  The best fit suggest a \oii\,\,interloper at the lowest redshift. The grey line is the same as \laea.
SED fittings for \otf LAE candidates. \otf (hollow circle), Subaru, HST, and Spitzer (filled circles for all three) fluxes are plotted along with SEDs calculated with HyperZ. (a) Candidate \laea: the solid line corresponds to the best fit SED, a $z=3.03$ starburst galaxy; the dashed line shows the fit at $z=1.45$ for an \oii\ interloper, and the gray line shows a scaled high--resolution version of the previous fit, obtained from a starburst galaxy template by \citep{calzetti:1994, kinney:1996}. %
(b) Candidate \laeb: the best fit corresponds to a $z\simeq5.4$ LBG. %
(c) Candidate \laec: the best fit is obtained for a $z=3.0$ (solid line) starburst galaxy; dashed and gray lines show the fit and scaled templates for a $z=1.45$ \oii\ interloper, as for \laea.
}\label{fig:LaeSeds}
    \end{figure} 

%% file: tableX.tex
 \begin{table}[t]
    \centering
    \caption{SED fits}\label{tab:interloper}
    %\small
    \begin{tabular}{c c c c}
    % HST-WFPC2 filter infomation at:
    %   http://www.stsci.edu/hst/wfpc2/documents/wfpc2_filters_archive.html
	\tableline \tableline \noalign{\smallskip} %
	ID	&	          \mc{Starburst}       &	\oii\tn{b}   \\
		&	Redshift\tn{a}	& $\chi^{2}$  &	$\chi^{2}$    \\
	\tableline \noalign{\smallskip} %
	\laea	&	3.0	&    26.2        &	31.3	\\
	\laeb	&	5.4	&    36.3        &	$\cdots$\\
	\laec	&	3.0	&    18.3        &	19.4	\\
	\dlla	&	5.4	&    33.1        &	$\cdots$\\
	\dllb	&	5.3	&    \pa 4.6     &	\pa 5.7	\\
	\lbga    &	2.4 &    10.1\tn{c}  &	$\cdots$\\
	\lbgb    &	5.5 &    3.1         &	$\cdots$\\
    \tableline
    \end{tabular}
      \tablenotetext{a}{SED fit done with Subaru, HST and Spitzer data.}
      \tablenotetext{b}{\oii\,\,candidates would be at $z\simeq1.45$.}
      \tablenotetext{c}{The shown best SED $\chi^{2}$ fit is for a young spiral galaxy.}
    \end{table}

%% file: fig10.tex
    \begin{figure}[t!]
      \vspace{-0.1in}
      \centering
      \includegraphics[width=\linewidth]{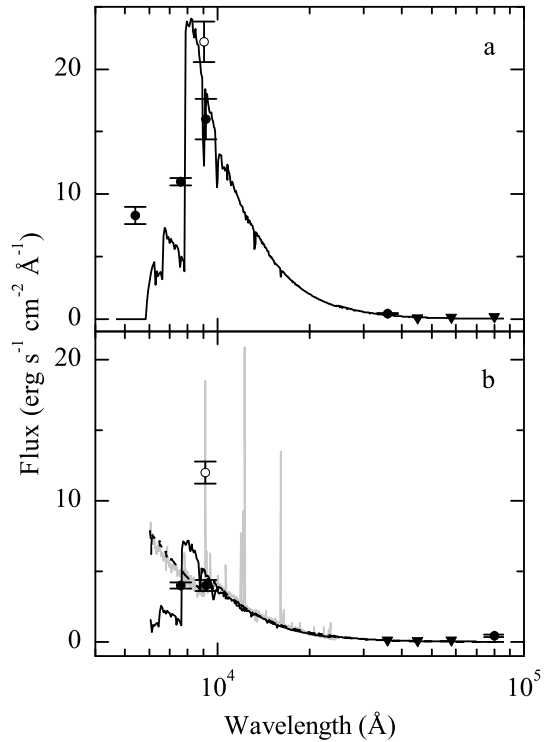}\\
      \caption{SED fittings for \otf LAE or LBG candidates. Fluxes obtained from \otf (hollow circle) and Subaru, HST, and Spitzer/IRAC (filled circles for the last three) images are plotted, along with SEDs fits calculated with HyperZ. (a) Candidate \dlla: the SED fit is for a starbusrst galaxy at redshift $z\simeq5.4$. The object is most likely and interloper. (b) Candidate \dllb: photometric data have been fitted at $z\simeq5.3$ (solid line) for a starburst galaxy, and at 1.45 (dashed line) for an \oii\ emitter. The best fit suggests a \oii\,\,interloper at the lowest redshift. The gray line is the same as in \laea in Figure \ref{fig:LaeSeds}.}\label{fig:LaeLbgSeds}
    \end{figure} 

%% file: fig11.tex
    \begin{figure}[t!]
      \centering
      \includegraphics[width=\linewidth]{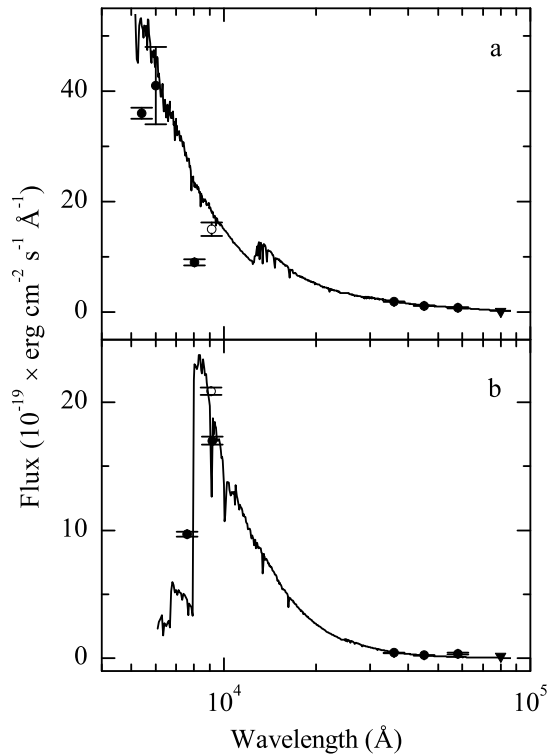}\\
      \caption{SED fittings for \otf LBG candidates. Fluxes obtained from \otf (hollow circles) and Subaru, HST, and Spitzer/IRAC (filled circles for the last three) images are plotted, along with SEDs calculated with HyperZ. (a) Candidate \lbga: the best SED fit is that of a spiral galaxy at redshift $z\simeq2.4$. (b) Candidate \lbgb: the photometric data are consistent with the SED of a starburst galaxy at $z=5.5$; The SED profile suggests the object may be a LBG at $z\gtrsim5.5$ or a high--redshift interloper.}\label{fig:LbgSeds}
    \end{figure}

    % Candidate \lbga: the SED's reddest break corresponds to the Balmer limit at 4000 {\AA}.
    % If this break were at the OTF-TF position (9123 {\AA}), z > (9123-4000)/4000=1.2808.
    % The second break corresponds to the Lyman break, starting from Lyman-alpha at 1216 {\AA}.
    % If this break were at the HST-F606W position (5997 {\AA}), z < (5997-1216)/1216=3.9317.
    % Thus, 1.3 < z < 4. 

%% file: tab3.tex
\begin{table*}[ht]
  \centering
  \caption{Simulation statistics}\label{tab:statistics}
  \begin{tabular}{llcrrr} %
  \tableline \tableline \noalign{\smallskip} %
  Instrument & Parameter\tn{a} & N\tn{b}  & Median & Q1\tn{c} & Q3\tn{d} \\
  \tableline \noalign{\smallskip} %
     IDEAL  & $z$			& 1290		& 6.55   & 6.53     & 6.58 \\
        	& {$L_{\la}$\tn{e}}    & $\cdots$ 	& 6.35   & 4.16     & 10.93 \\
 		& {$EW_0$\tn{f}}      & $\cdots$ 	& 94      & 56        & 175 \\
		& {$FWHM_0$\tn{f}} & $\cdots$ 	& 1.63   & 1.25     & 2.13 \\
     Subaru  & $z$             	& 1108     	& 6.54   & 6.51     & 6.56 \\
		& {$L_{\la}$\tn{e}}    & $\cdots$ 	& 7.49   & 4.85     & 13.23 \\
		& {$EW_0$\tn{f}}      & $\cdots$ 	& 105    & 66        & 187 \\
		& {$FWHM_0$\tn{f}} & $\cdots$ 	& 1.65   & 1.26     & 2.19 \\
     \otf    & $z$            	 	& 2708     	& 6.55   & 6.51     & 6.59 \\
		& {$L_{\la}$\tn{e}}    & $\cdots$ 	& 6.00   & 3.95     & 11.14 \\
		& {$EW_0$\tn{f}}      & $\cdots$ 	& 67      & 34        & 140 \\
		& {$FWHM_0$\tn{f}} & $\cdots$ 	& 1.67   & 1.27     & 2.16 \\
     \tableline
  \end{tabular}
  % Column~1 identifies the instrument. Column~2 designates the modeled parameters studied (redshift, luminosity of the \la line, and the rest-frame EW and FWHM). Column~3 indicates the number of detections over the 5000 simulations. Columns~4, 5 and 6 correspond to the median, the first and the third quartiles of the parameter distribution of the candidates, respectively.
  \tablenotetext{a}{Modeled parameters studied (redshift, luminosity of the \la line, and the rest-frame EW and FWHM).}
  \tablenotetext{b}{Number of detections over the 5000 simulations.}
  \tablenotetext{c}{First quartile of the parameter distribution of the recovered LAEs.}
  \tablenotetext{d}{Idem for the third quartile.}
  \tablenotetext{e}{In units of $10^{43}\,\ergs$.}
  \tablenotetext{f}{In {\AA}.}
\end{table*}

%% file: fig13.tex
    \begin{figure}[]
      \includegraphics[width=\linewidth]{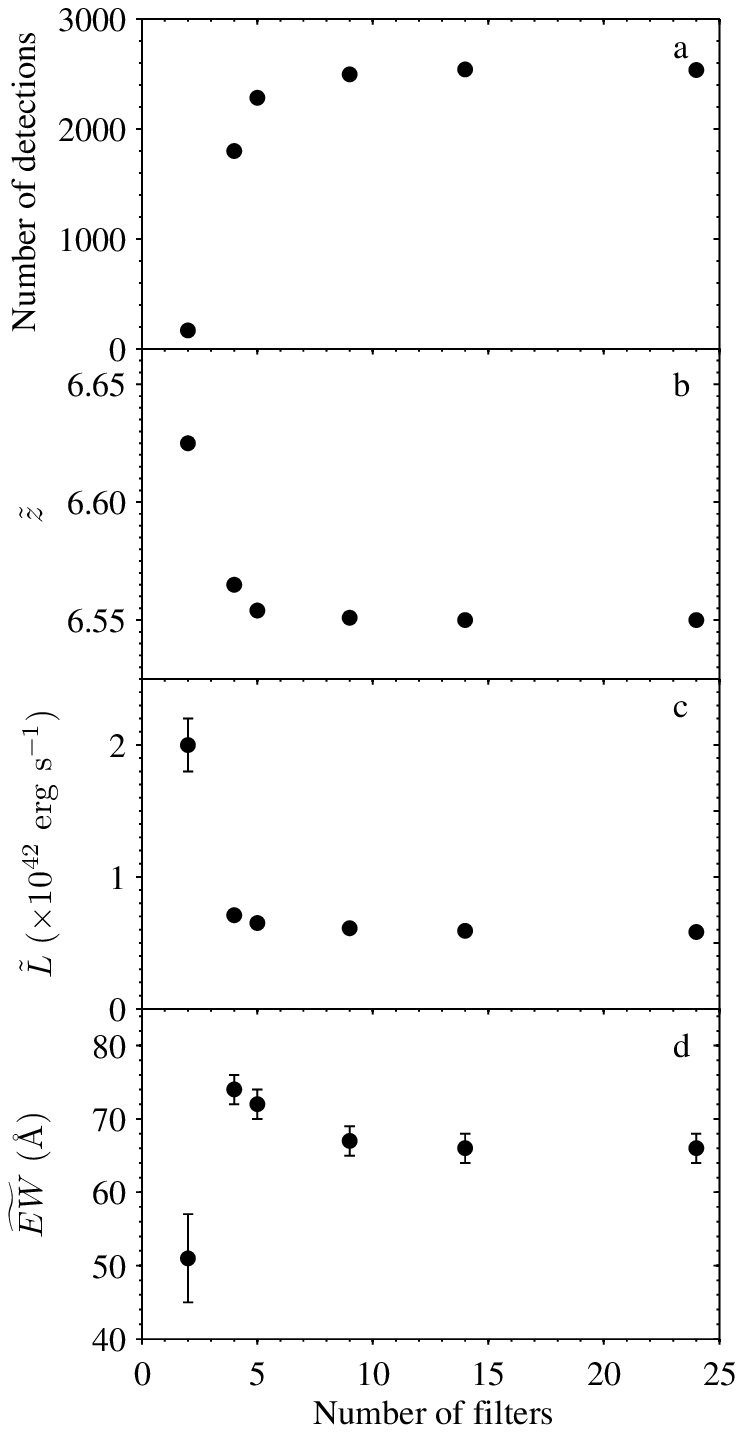}\\
      \caption{Dependences on the number of filters. This figure shows the results obtained using different sets of ideal (rectangular) filters on the LAE simulation dataset.  Each of these sets consist of different number of filters, but they cover the same wavelength range than the original \otf set.  Filters in a given set share the same FWHM, and their centers differ in steps of half this FWHM.  For detections, we have adopted the same criteria as in the case of \otf.  Panels show (a) the number of detections, (b) the median redshift, (c) the line luminosity, and (d) the restframe equivalent width. The results for 7 or more filters are similar, but for smaller number of  filters the number of detections falls dramatically, and the other parameters also change significantly.} \label{fig:filtnumdep}
    \end{figure}

%% file: tab5.tex
\begin{table}[t]
  \centering
  \caption{Dependence of detections on the number of filters.} \label{tab:nfilt}
  \begin{tabular}{rrrr@{$\pm$}l}
		\tableline \tableline \noalign{\smallskip} %
		\mca{FWHM}	& \mca{Number}& \mr{Detections}	& \mc{EW$_0$\tn{a}} \\
		\mca{(\AA)}	& \mca{of filters}& 				& \mc{(\AA)} \\
    \tableline \noalign{\smallskip}
		12	&	24	&	2535	&	66	&	2 \\
		20	&	14	&	2542	&	66	&	2 \\
		30	&	 9	&	2497	&	67	&	2 \\
		50	&	 5	&	2282	&	72	&	2 \\
		60	&	 4	&	1799	&	74	&	2 \\
		100	&	 2	&	 168	&	51	&	6 \\
    \tableline
	\end{tabular}
	\tablenotetext{a}{Median and error. The errors ($e$) are computed from the \\
				   interquartile range($r_q$) by $e = 0.7413 r_q N^{-1/2}$, where \\ 
				   $N$ is the number of detections.}
\end{table}

%% file: tab4.tex
\begin{table}[t]
  \centering
  \caption{\otf Survey Expected Quantities}\label{tab:survey}
  \begin{tabular}{@{}lrr@{}}
    \tableline \tableline \noalign{\smallskip} %
    & \mc{Observations\tn{a}} \\
    & Planed & Accomp. \\
    \tableline \noalign{\smallskip} %
    Slices\tn{b}						&     24		&       5     \\
    Irradiance limit\tn{c}				&       4		&       9     \\
    L$_{\la}$ limit\tn{d}				&       2.1		&       4.7  \\
    Volume LAEs\tn{e}				& 8760		& 1503     \\
    Expected LAEs\tn{f} 				&       4.2		&       0.1  \\
%   Idem with 20\% magnification		&       5.7		&       0.2  \\
%   Idem cosmic variance			 	&    2$\pm$3	&    0$\pm$1 \\
    L$_{\mathrm{[O\,{II}]}}$ limit\tn{g}	&       2.9		&       6.4  \\
    Volume \oii\tn{e}					&    122		&     21     \\
    Expected \oii\ (Takahashi)\tn{h} 		&       7.4	 	&       0.8  \\
%   Idem with 10\% magnification		&       7.8		&       0.9  \\
%   Idem cosmic variance				& 3$\pm$3	&    0$\pm$1 \\
    Expected \oii\ (Dressler)\tn{i} 				&       2.5	 	&       0.2  \\
%   Idem with 10\% magnification		&       2.6		&       0.2  \\
%   Idem cosmic variance				& 1$\pm$2	&    0$\pm$1 \\
   \tableline
  \end{tabular}
  \tablenotetext{a}{State of the observations: Planned or Accomplished.}
  \tablenotetext{b}{Number of wavelength slices.}
  \tablenotetext{c}{Irradiance lower limit ($\times 10^{-18} \, \ergscm$).}
  \tablenotetext{d}{\la luminosity lower limit ($\times 10^{42} \, \ergs$).}
  \tablenotetext{e}{Proper volume covered (Mpc$^{-3}$).}
  \tablenotetext{f}{Number of expected LAEs obtained through the Schechter \\
    function with \citet{kashikawa:2011} parameters. Numbers \\ 
    are given with a precission of one  decimal place, rather than \\
    integers, to ensure that at least one significant digit is shown.}
  \tablenotetext{g}{\oii\ luminosity lower limit ($\times 10^{40} \, \ergs$).}
  \tablenotetext{h}{\citet{takahashi:2007}}.
  \tablenotetext{i}{\citet{dressler:2011}}.
\end{table}

% Cosmic variance calculated using the Cosmic Variance Calculator:
% 	http://casa.colorado.edu/~trenti/CosmicVariance.html,
% 	Trenti & Stiavelli (2008), ApJ, 676, 767 
% with parameters:
% 	- Halo filling factor: 0.2 for LAEs, 0.5 for [OII] interlopers
% 	- Completeness: 0.5 for LAEs and [OII] interlopers.

%% file: tab6.tex
\begin{table}[t]
  \centering
  \caption{Observer-frame FWHM for $z\simeq 6.5$ LAEs and interlopers.}\label{tab:linefwhms}
  \begin{tabular}{lrrr}
		\tableline \tableline \noalign{\smallskip} %
		\mr{Line}	& \mr{Redshift}	& \mca{Velocity}		& \mca{FWHM}	\\
				&				& \mca{(km s$^{-1}$)}	& \mca{(\AA)} \\
    \tableline \noalign{\smallskip}
		$\la_{\lambda 1216}$  & 6.50 & 400 & 12.16 \\
		\oii$_{\lambda\lambda 3726-9}$ & 1.45 & -- & $\sim10\phantom{.00}$ \\
%		\mbox{[O III]$_{\lambda 3726}$} & 1.45 & 100 & 3.04 \\
		\mbox{[\ion{O}{3}]$_{\lambda 5007}$ } & 0.82 & 100 & 3.04 \\
		$\rm{H\alpha_{\lambda 6563}}$ & 0.39 & 100 & 3.04 \\
    \tableline
	\end{tabular}
\end{table}

%% file: fig12.tex
\begin{figure}[t]
  \includegraphics[width=\linewidth]{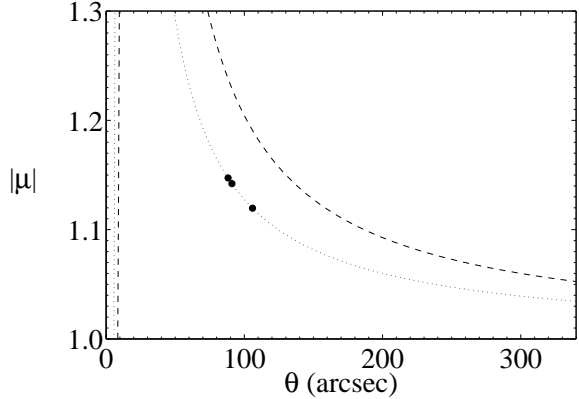}\\
  \caption{Magnification as a function of the angular separation to the center of the cluster of galaxies. The dotted line shows the absolute value of the magnification $\mu$ as a function of the angular separation $\theta$ for the cluster \cluster, using a singular isothermal sphere model. Filled circles correspond to \oii\ interloper candidates.}\label{fig:magnification}
\end{figure} 